%% file: mainjournal.tex
\RequirePackage{fix-cm}
\documentclass{svjour3}                     
\smartqed  
\usepackage{graphicx}

\usepackage{headers}
\usepackage{macro}

\begin{document}

\title{Spiking Neural Networks modelled as Timed Automata
}
\subtitle{with parameter learning}


\author{Elisabetta De Maria         \and
        Cinzia Di Giusto \and
        Laetitia Laversa        
}


\institute{Universit\'{e} C\^{o}te d'Azur, CNRS, I3S, France}

\date{Received: date / Accepted: date}

\maketitle

\begin{abstract}

In this paper we present a novel approach to automatically infer parameters of \SNN{s}.
	Neurons are modelled as timed automata waiting for inputs on a number of different channels (synap-ses), for a given amount of time (the accumulation period). When this period is over, the current \emph{potential} value is computed  considering current and past inputs. 
If this potential overcomes a given \emph{threshold}, the automaton emits a broadcast signal over its output channel, otherwise it restarts another accumulation period.
	After each emission, the automaton remains inactive for a fixed \emph{refractory period}.
	Spiking neural networks are formalised as sets of automata, one for each neuron, running in parallel and sharing channels according to the network structure.
Such a model is formally validated against some crucial properties defined via proper temporal logic formulae.
 The model is then exploited to find an assignment for the synaptical weights of neural networks such that they can reproduce a given behaviour. The core of this approach consists in identifying  some correcting actions adjusting synaptical weights and back-propagating them until the expected behaviour is  displayed. A concrete case study is discussed.
 \keywords{Neural Networks\and Parameter Learning \and Timed Automata, Temporal Logic \and Model Checking.}
\end{abstract}

\section{Introduction} \label{sec:introduction}
\input{introduction.tex}

\section{Preliminaries} \label{sec:preliminaries}
\input{preliminaries.tex}

\section{Leaky Integrate and Fire Model and Mapping to Timed Automata} \label{sec:lif}
	\input{neural_networks.tex}

	\input{encoding.tex}

\section{Validation of the  model} \label{sec:valid}
		\input{inner.tex}

		\input{n_props.tex}

\section{{Parameter inference }}\label{sec:learning}
	\input{learning.tex}


\section{Related Work} \label{sec:related}
\input{related.tex}

\section{{Conclusion}}\label{sec:conclusion}
	\input{conclusion.tex}

\begin{acknowledgements}
We are grateful to Giovanni Ciatto for his preliminary implementation work and for his enthusiasm in collaborating with us.
\end{acknowledgements}

\bibliographystyle{spmpsci}      

\bibliography{bibliography}   

\end{document}

%% file: introduction.tex

The brain behaviour is the object of thorough  studies: researchers are interested not only in the inner functioning of neurons (which are its elementary components), their interactions and the way these aspects participate to the ability to move, learn or remember, typical of living beings; but also in reproducing such capabilities (emulating nature), \eg within robot controllers, speech/text/face recognition applications, etc.
In order to achieve a detailed understanding of the brain functioning, both neurons behaviour and their interactions must be studied.
Several models of the neuron behaviour have been proposed: some of them make neurons behave as binary threshold gates, other ones exploit a sigmoidal transfer function, while, in many cases, differential equations are employed.
According to \cite{paugam-moisy2012,maass97}, three different and progressive \emph{generations} of neural networks can be recognised: \begin{enumerate*}[label=(\roman*)]
	\item \emph{first generation} models handle discrete inputs and outputs and their computational units are threshold-based transfer functions; they include McCulloch and Pitt's threshold gate model \cite{mcculloch1943}, the perceptron model\cite{freund1999}, Hopfield networks \cite{hopfield88}, and Boltzmann machines \cite{ackley88};
	\item \emph{second generation} models exploit real valued activation functions, \eg the sigmoid function, accepting and producing real values: a well known example is the multi-layer perceptron \cite{cybenko89,rumelhart88};
	\item \emph{third generation} networks are known as \SNN{s}. They extend second generation models treating time-dependent and real valued signals often composed by \emph{spike trains}. Neurons may fire output spikes according to threshold-based rules which take into account input spike magnitudes and occurrence times \cite{paugam-moisy2012}.
\end{enumerate*}

The core of our analysis are \emph{\SNN{s}} \cite{Gerstner2002}. Because of the introduction of timing aspects they are considered closer to the actual brain functioning than other generations models.
Spiking neurons  emit spikes taking into account input impulses strength and their occurrence instants.
Models of this sort are of great interest, not only because they are closer to natural  neural networks behaviour, but also because the temporal dimension allows to represent information according to various \emph{coding schemes} \cite{recce99,paugam-moisy2012}:  \eg the a\-mount of spikes occurred within a given time window (\emph{rate} coding), the reception/absence of spikes over different synapses (\emph{binary} coding), the relative order of spikes occurrences  (\emph{rate rank} coding), or the precise time difference between any two successive spikes (\emph{timing} coding).

Several spiking neuron models have been proposed in the literature, having different complexities and capabilities.
In \cite{izhikevich04}, Izhikevich classifies spiking neuron models according to some \emph{behaviour} (\ie typical responses to an input pattern) that they should exhibit in order to be considered biologically relevant.
The \LIandF  (\lif) model \cite{lapicque1907}, where past inputs relevance exponentially decays with time, is one of the most studied neuron models because  it is straightforward  and  easy to use \cite{izhikevich04,paugam-moisy2012}. On the other end of the spectrum, the Hodgkin-Huxley (H-H) model \cite{hodgkin52} is one of the most complex being composed by four differential equations comparing neurons to electrical circuits.
In \cite{izhikevich04}, the H-H model can reproduce all behaviours under consideration, but the simulation process is really expensive even for just a few neurons being simulated for a small amount of time.
Our aim is to produce a neuron model being meaningful from a biological point of view but also amenable to formal analysis and verification, that could be therefore used to detect non-active portions within some network (\ie the subset of neurons not contributing to the network outcome), to test whether a particular output sequence can be produced or not, to prove that a network may never be able to emit, to assess if a change to the network structure can alter its behaviour, or to investigate (new) learning algorithms which take time into account.

In this paper we focus on the \emph{\LIandF} (LI\&F) model originally proposed in \cite{lapicque1907}.  It is a computationally efficient
approximation of single-compartment model \cite{izhikevich04} and is abstracted enough to be able to apply formal verification techniques such as model-checking. Here we  work on an extended version of the discretised formulation proposed in \cite{demaria16}, which relies on the notion of logical time. Time is considered
as a sequence of logical discrete instants, and an instant is a point in time where
external input events can be observed, computations can be done, and outputs can be emitted.
The variant we introduce here takes into account some new time-related aspects, such as a lapse of time
in which the neuron is not active, i.e., it cannot receive and emit. We encode \lif  networks into \TAs: we show how to define the behaviour of a single neuron and how to build a network of neurons.
Timed automata \cite{alur94} are finite state automata extended with timed behaviours: constraints are allowed to limit the amount of time an automaton can remain within a particular state, or the time interval during which a particular transition may be enabled. Timed automata networks are sets of automata that can synchronise over \emph{channel} communications.

Our modelling of \SNN{s} consists of \TAN{s} where each neuron is an automaton. Its behaviour consists in  accumulating the weighted sum of inputs, provided by a number of ingoing weighted synapses, for a given amount of time. Then, if  the \emph{potential} accumulated during the last and previous accumulation periods overcomes a given threshold, the neuron fires an output over the outgoing synapse. Synapses are channels shared between the timed automata representing neurons, while \emph{spike} emissions are represented by \emph{broadcast synchronisations} occurring over such channels. Timed automata are also exploited to produce or \emph{recognise} precisely defined spike sequences.

As a first main contribution, we analyse some intrinsic properties of the proposed
model, e.g., the maximum threshold value allowing a neuron to
emit, or the lack of inter-spike memory, preventing the behaviour of a neuron
from being influenced by what happened before the last spike.
Furthermore, we encode in temporal logics all the behaviours (or capabilities) a \lif model should be able to
reproduce according to Izhikevich and we exploit model checking to prove these behaviours are reproducible in our model. Izhikevich
also identifies a set of behaviours which are not expected to be reproducible by
any \lif model. We prove these
limits to hold for our model, too, and we provide, for each non-reproducible behaviour, an extension of the model allowing to reproduce it.

As a second main contribution, we exploit our au\-tom\-a\-ta-based modelling to propose a new methodology for parameter inference  in \SNN{s}. In particular, our approach allows to find an assignment for the synaptical weights of a given neural network such that it can reproduce a given behaviour.
We apply the proposed approach to find suitable parameters in mutual inhibition networks, a well studied class of
networks in which the constituent neurons inhibit each other neuron's activity \cite{M87BC}.

This paper is an extended and revised version of the conference papers \cite{bioinfo18} and \cite{csbio17}. In particular, in section \ref{sec:learning} we propose a refined version of the \ABP algorithm. Furthermore, we add a second learning technique that is based on simulation instead of model checking. 

The rest of the paper is organised as follows: in Section \ref{sec:preliminaries} we recall definitions of \TAN{s}, temporal logics, and model checking; in Section \ref{sec:lif} we describe our reference model, the \lif one, and its encoding into \TAN{s}; in Section \ref{sec:valid} we study some intrinsic properties of the obtained model and we validate it against its
ability of reproducing or not some behaviours;
in Section \ref{sec:learning} we develop the novel parameter learning approach and we introduce a case study; in Section \ref{sec:related} we give an overview of the related work.
Finally, Section \ref{sec:conclusion} summarises our contribution and presents some future research directions.

%% file: preliminaries.tex

In this section we introduce the formal tools we adopt in the rest of the paper, namely \TAs{} and temporal logics.

\subsection{Timed Automata.}
Timed automata \cite{alur94} are a powerful theoretical formalism for modelling and verifying real time systems.  A timed automaton is an annotated directed (and connected) graph,
with an initial node and provided with a finite set of non-negative real
variables called \emph{clocks}.
Nodes (called \emph{locations}) are annotated with \emph{invariants}
(predicates allowing to enter or stay in a location), arcs  with \emph{guards},
\emph{communication labels}, and possibly with some variables upgrades and clock \emph{resets}.
Guards are conjunctions of elementary predicates of the form
$x~\op~c$, where $\op\in\{>,\geq,=,<,\leq\}$, $x$ is a clock, and
$c$ a (possibly parameterised) positive integer constant.
As usual, the empty conjunction is interpreted as true.
The set of all guards and invariant predicates will be denoted by $G$.


\begin{definition}\label{def:ta} A \emph{timed automaton} $\TA$
is a tuple $(L,l^0,X,\\\Sigma,\arcs,\inv)$, where
\begin{itemize}
	\item $L$  is a set of locations with $l^0\in L$ the initial one
    \item $X$ is the set of clocks,
	\item $\Sigma$ is a set of communication labels,
	\item $\arcs \subseteq L \times (G \cup \Sigma \cup U) \times L$
is a set of arcs between locations with a guard in $G$,
a communication label in $\Sigma\cup\{\varepsilon\}$, and a set of variable upgrades (\eg clock resets);
  \item $\inv: L \rightarrow G$ assigns invariants to locations.
\end{itemize}
\end{definition}

It is possible to define a synchronised product of a set of timed automata
that work and synchronise in parallel.
The automata are required to have disjoint sets of locations, but
may share clocks and communication labels which are used for synchronisation.
We restrict communications to be  \emph{broadcast}  through labels $b!,b?\in\Sigma$, meaning that
	a set of automata can synchronise if one is emitting;
	notice that a process can always emit (\eg $b!$)
	and the receivers ($b?$) must synchronise if they can.
	
Locations can be normal, urgent or committed.  Urgent locations force the time to freeze, committed ones freeze time and the automaton must leave the location as soon as possible, \ie they have higher priority.

The synchronous product $\TA_1\parallel \ldots \parallel \TA_n$
of timed automata, where 
$\TA_j = (L_j, l^0_j,X_j,\Sigma_j,\arcs_j,\inv_j)$  and  $L_j$ are
pairwise disjoint sets of locations for each $j\in[1,\ldots,n]$, is
the timed automaton $$\TA=(L, l^0,X,\Sigma,\arcs,\inv)$$ such that:
\begin{itemize}
\item $L=L_1\times\ldots\times L_n$ and $l^0=(l^0_1,\ldots,l^0_n)$,
$X=\bigcup_{j=1}^n X_j$, $\Sigma=\bigcup_{j=1}^n \Sigma_j$,
\item $\forall l=(l_1, \ldots, l_n)\in L\colon \inv(l) = \bigwedge_j \inv_j(l_j)$,
\item $\arcs$ is the set of arcs
$(l_1, \ldots, l_n) \stackrel{g,a,r}{\longrightarrow} (l'_1, \ldots, l'_n)$
such that
	for all $1\leq j \leq n$  then $l_j'=l_j$.
\end{itemize}

Its semantics is the one of the underlying timed automaton $\TA$ with the following notations.
A location is a vector $l = (l_1, \ldots, l_n)$.
We write $l[l'_j/l_j, j\in S]$ to denote the location $l$ in which the
$j^{th}$ element $l_j$ is replaced by $l'_j$, for all $j$ in some set $S$.
A valuation is a function $\nu$ from the set of clocks to the non-negative reals.
Let $\mathbb{V}$ be the set of all clock valuations, and $\nu_0(x) = 0$
for all $x \in X$.
We shall denote by $\nu\vDash F$ the fact that the valuation $\nu$
satisfies (makes true) the formula $F$.
If $r$ is a clock reset, we shall denote by $\nu[r]$
the valuation obtained after applying the clock reset $r\subseteq X$ to $\nu$;
and if $d\in\mathbb{R}_{> 0}$ is a delay,
$\nu+d$ is the valuation such that, for any clock $x\in X$,
$(\nu+d)(x)=\nu(x)+d$.

The semantics of a synchronous product $\TA_1\parallel \ldots \parallel \TA_n$
is defined as a timed transition system $(S,s_0,\rightarrow)$,
where $S = (L_1 \times,\ldots\times L_n) \times \mathbb{V}$ is the set of states,
$s_0 = (l^0, \nu_0)$ is the initial state, and
$\rightarrow \subseteq S \times S$ is the transition relation defined by:
\begin{itemize}	
	\item (silent): $(l,\nu) \rightarrow (l',\nu')$
	if there exists $l_i \stackrel{g,\varepsilon,r}{\longrightarrow} l'_i$, for some $i$,
	such that $l'=l[l'_i/l_i]$, $\nu\vDash g$
	and $\nu'=\nu[r]$,

	\item (broadcast): $(\bar{l},\nu) \rightarrow (\bar{l'},\nu')$ if
	there exists an output arc $l_j \stackrel{g_j,b!,r_j}{\longrightarrow} l_j'\in \arcs_j$
	and a (possibly empty) set of input arcs of the form
	$l_k \stackrel{g_k,b?,r_k}{\longrightarrow} l_k'\in \arcs_k$ such that
	for all $k\in K=\{k_1,\ldots,k_m\}\subseteq\{l_1,\ldots,l_n\}\setminus\{l_j\}$,
	the size of $K$ is maximal, $\nu\vDash \bigwedge_{k\in K\cup\{j\}} g_k$,
	$l'=l[l'_k/l_k, k\in K\cup\{j\}]$ and $\nu'=\nu[r_k, k\in K\cup\{j\}]$;
	
	\item (timed): $(l,\nu)\rightarrow (l,\nu+d)$ if $\nu+d\vDash \inv(l)$.
\end{itemize}

The valuation function $\nu$ is extended to handle a set of shared bounded integer variables: predicates concerning such variables can be part of edges guards or locations invariants, moreover variables can be updated on edges firings but they cannot be assigned to or from clocks.
\begin{example}

In \figref{fig:ta} we
consider the network of timed automata $\TA_1$ and $\TA_2$ with broadcast  communications, and we give a possible run.
$TA_1$ and $TA_2$ start in the $l_1$ and $l_3$ locations, respectively, so the initial state is $[(l_1,\, l_3);\ x = 0]$.
	A \emph{timed} transition produces a delay of 1 time unit, making the system move to state $[(l_1,\, l_3);\ x = 1]$.
	A \emph{broadcast} transition is now enabled, making the system move to state $[(l_2,\, l_3);\ x = 0]$, broadcasting over channel $a$ and resetting the $x$ clock.
	Two successive  \emph{timed} transitions (0.5 time units) followed by a \emph{broadcast} one will eventually lead the system to state $[(l_2,\, l_4);\ x = 1]$.
\hfill $\diamond$
\begin{figure}[t]
\centering
\subfigure[The timed automata network $\TA_1\parallel \TA_2$.] 
{ \label{product}

\begin{tikzpicture}
	\node(ta1) at (-1.5,0){\textbf{TA$_1$}};
	\node[draw,circle] (A) at (0,0) {$l_1$} ;
	\node[draw, circle] (B) at (3,0){$l_2$};
	\node[draw, circle, minimum width = 0.6cm](A2) at (0,0) {};

	\draw(0,-0.5) node[left, scale = 0.8] {\textbf{$x < 2$}};	
	\draw(3,-0.5) node[left, scale = 0.8] {\textbf{$x < 2$}};			
	\draw[->, >=latex] (A) -- (B);
	\draw(0.7,0.8) node[right, scale = 0.8] {$G : x = 1$}; 
	\draw(0.7,0.5) node[right, scale = 0.8] {$S : a! $};
	\draw(0.7, 0.2) node[right, scale = 0.8] {$U : x := 0$};  
	\draw[->, >=latex] (B) to[out= 45, in =90] (4, 0) to[out = 270, in = -45] (B);
 	\draw(4.3, 0.4) node[right, scale = 0.8] {$G : x > 0$};
 	\draw(4.3, 0.1) node[right, scale = 0.8] {$S : b! $}; 
 	\draw(4.3, -0.2) node[right, scale = 0.8] {$U : - $};

 	\node(ta2) at (4.5,-2){\textbf{TA$_2$}};
	\node[draw,circle] (C) at (0,-2) {$l_3$} ;
	\node[draw, circle] (D) at (3,-2){$l_4$};
	\node[draw, circle, minimum width = 0.6cm](C2) at (0,-2) {};

	\draw[->, >=latex] (C) -- (D);
	\draw(0.7,-2.2) node[right, scale = 0.8] {$G : x = 1$}; 
	\draw(0.7,-2.5) node[right, scale = 0.8] {$S : - $};
	\draw(0.7, -2.8) node[right, scale = 0.8] {$U : - $};  
	\draw[->, >=latex] (C) to[out= 135, in =90] (-1, -2) to[out = 270, in = 225] (C);
 	\draw(-2.2, -1.8) node[right, scale = 0.8] {$G : \true $};
 	\draw(-2.2, -2.1) node[right, scale = 0.8] {$S : a? $}; 
 	\draw(-2.2, -2.4) node[right, scale = 0.8] {$U : x:=0 $}; 
 	
\end{tikzpicture} 
%
}
\quad
\subfigure[A possible run.]{\small
$
\begin{array}{c}
\left[(l_1,\, l_3);\ x = 0\right] \\ \downarrow \\
\left[(l_1,\, l_3);\ x = 1\right] \\ \downarrow \\
\left[(l_2,\, l_3);\ x = 0\right] \\ \downarrow \\
\left[(l_2,\, l_3);\ x = 0.5\right] \\ \downarrow \\
\left[(l_2,\, l_3);\ x = 1\right] \\ \downarrow \\
\left[(l_2,\, l_4);\ x = 1\right]
\end{array}
$
}
\caption{
	A network of timed automata with a possible run.
}
\label{fig:ta}
\end{figure}
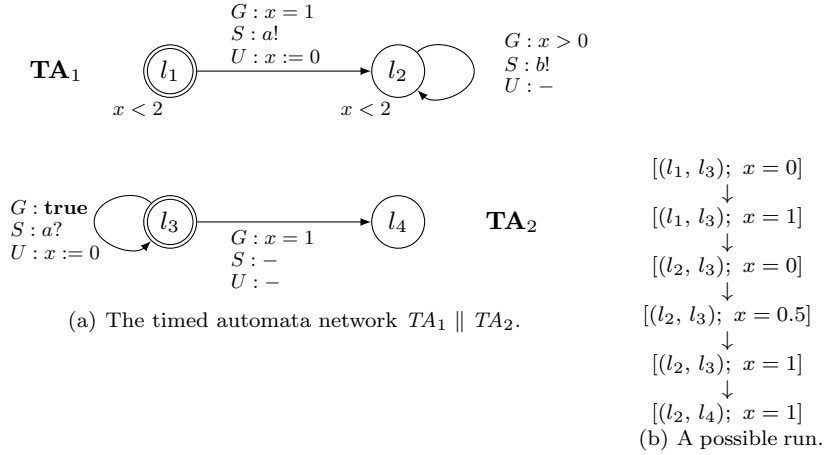
\end{example}

Throughout our modelling, we have used the specification and analysis tool \Uppaal \cite{bllpw:dimacs95}, which provides the possibility of designing and simulating \TAN{s} on top of the ability of testing networks against temporal logic formulae. All figures depicting timed automata follow the graphic conventions of the tool (\eg initial states are denoted with a double circle).

\subsection{ Temporal Logics and Model Checking}
Model checking is one of the most common approaches to the verification of software and hardware (distributed) systems \cite{C99mit}.
It allows to automatically prove whether a system verifies or not a given specification.
In order to apply such a technique, the system at issue should be encoded as a finite transition
system and the specification should be written using propositional temporal logic.
Formally, a transition system over a set $AP$ of atomic propositions is a tuple $M=(Q,T,L)$,
where $Q$ is a finite set of states, $T \subseteq Q \times Q$ is a total transition relation, 
and $L: Q \rightarrow 2^{AP}$ is a labelling function that maps every state
into the set of atomic propositions that hold at that state.

Temporal formulae  describe the dynamical evolution of a given system.
The computation tree logic
CTL$^\ast$ allows to describe properties of computation trees. Its formulas are obtained by
(repeatedly) applying boolean connectives ($\wedge$, $\vee$, $\neg$, $\rightarrow$), \emph{path quantifiers},
and \emph{state quantifiers} to atomic formulas. The path quantifier
$\textbf{A}$ (resp., $\textbf{E}$) can be used to state that all the paths
(resp., some path) starting from a given state have some property.
The state quantifiers are
$\textbf{X}$  (next time),
 which specifies that a property holds at the next state of a path,
$\textbf{F}$ (sometimes in the future),
which requires a property to hold at some state on the path,
$\textbf{G}$ (always in the future),
 which imposes that a property is true at every state on the path,
and $\textbf{U}$ (until),
which holds if there is a state on the path
where the second of its argument properties holds and, at every
preceding state on the path, the first of its two argument
 properties holds. Given two formulas $\varphi_1$ and $\varphi_2$,
 in the rest of the paper we use the shortcut $\varphi_1 \rightsquigarrow \varphi_2$ to denote the
 liveness property $AG(\varphi_1 \rightarrow AF \varphi_2)$, which can be read as \textquotedblleft $\varphi_1$ always leads to $\varphi_2$ \textquotedblright. 

The branching time logic CTL is a fragment of CTL$^\ast$
that allows quantification over the paths starting from a given
state. Unlike CTL$^\ast$, it constrains every state quantifier to be
immediately preceded by a path quantifier.

Given a transition system $M=(Q,T,L)$, a state $q \in
Q$, and a temporal logic formula $\varphi$ expressing some desirable
property of the system, the \emph{model checking problem} consists
of establishing whether $\varphi$ holds at $q$ or not, namely,
whether $M, q \models \varphi$.

%% file: neural_networks.tex

Spiking neural networks \cite{maass97} are modelled as directed weigh\-ted graphs where vertices are computational units and edges represent \emph{synapses}.
		The signals propagating over synapses are trains of impulses: \emph{spikes}.
		Synapses may modulate these signals according to their weight: 
		 \emph{excitatory} if positive, or \emph{inhibitory} if negative.	
		
		The dynamics of neurons is governed by their \emph{membrane potential} (or, simply, \emph{potential}), representing the difference of electrical potential across the cell membrane.
		The membrane potential of each neuron depends on the spikes received over the ingoing synapses. 
		Both current and past spikes are taken into account, even if old spikes contribution is lower.
		 In particular, the \emph{leak factor} is  a measure of the neuron memory about past spikes.
		The neuron outcome is controlled by the algebraic difference between its membrane potential and its \emph{firing threshold}: it is enabled to fire (\ie  emit an output impulse over \emph{all} outgoing synapses) only if such a difference is non-negative. 	Spike propagation is assumed to be instantaneous.
		Immediately after each emission the neuron membrane potential is reset and the neuron stays in a \emph{refractory period} for a given amount of time. During this period it has no dynamics: it cannot increase its potential as any received spike is lost and therefore it cannot emit any spike.

\begin{definition}[Spiking Integrate and Fire Neural Network]
			\label{def.snn}
			\emph{A spiking integrate and fire neural network} is a tuple $(V,\, A,\, w)$, where: \begin{itemize}
				\item $V$ are spiking integrate and fire neurons,
				\item $A \subseteq V \times V$ are synapses,
				\item $w: A \rightarrow \mathbb{Q} \cap [-1,1]$ is the synapse weight function associating to each synapse $(u,\, v)$ a weight $w_{u,v}$.
			\end{itemize}
            We distinguish three disjoint sets  of neurons:  $V_{i}$ (input neurons), $V_{int} $ (intermediary neurons), and  $V_{o}$ (output neurons), with $V=V_{i} \cup V_{int} \cup V_{o}$.\\
            \emph{A spiking integrate and fire neuron} $v$ is characterized by a parameter tuple $$(\theta_v,\tau_v,\lambda_v, p_v, y_v),$$ where:
            \begin{itemize}
				\item $\theta_v \in \mathbb{N}$ is the \emph{firing threshold},
                \item $\tau_v \in \mathbb{N}^+$ is the \emph{refractory} period,
                \item $\lambda_v \in \mathbb{Q} \cap [0, 1]$ is the \emph{leak factor}.
            \end{itemize}
            The dynamics of a \emph{spiking integrate and fire neuron} $v$ is given by:
                \begin{itemize}
				\item $p_v: \mathbb{N} \rightarrow \mathbb{Q}^+_0$ is the [membrane] \emph{potential} function defined as
                \[
p_v(t) =\left\{ \begin{array}{l}
\sum_{i=1}^{m} w_i \cdot x_i(t), \quad \text{if }p_v (t-1) \geqslant \theta_v\\
\sum_{i=1}^{m} w_i \cdot x_i(t) + \lambda_v \cdot p_v(t - 1), \text{ otherwise.}
\end{array}\right.
\]
		with $p_v(0)=0$ and where  $x_i(t) \in \{0, 1\}$ is the signal received at the time $t$ by the neuron through its $i^{th}$ out of $m$ input synapses (observe that the past potential is multiplied by the leak factor while current inputs are not weakened),
                \item $y_v: \mathbb{N} \rightarrow \{0, 1\}$ is the neuron output function,  defined as
                \begin{equation*}
			y_v(t) = \begin{cases}
				1 & \mathrm{if}\  p_v(t) \geqslant \theta_v \\
				0 & \mathrm{otherwise.}
			\end{cases}
		\end{equation*}
			\end{itemize}	
		\end{definition}


As shown in the previous definition, the set of neurons of a spiking integrate and fire neural network can be classified into input, intermediary, and output ones. 
Each input neuron can only receive as input external signals (and not other neurons' output).
The output of each output neuron is considered as an output for the network. Output neurons are the only ones whose output is not connected to other neurons.

%% file: encoding.tex

		
		We present here our modelling of spiking integrate and fire neural networks (in the following denoted as neural networks) via \TAN{s}. 
Let $S=(V,A,w)$ be a neural network, $G$ be a set of \emph{input generator} neurons (these fictitious neurons are connected to input neurons and generate input sequences for the network), and $O$ be a set of \emph{output consumer} neurons (these fictitious neurons are connected to the broadcast channel of each output neuron and  aim at consuming their emitted spikes). The corresponding \TAN is obtained as the synchronous product of the encoding of input generator neurons, the neurons of the network (referred as standard neurons in the following), and output consumers neurons.
		More formally:
		\[
			\enc{S} = (\bigpar_{n_g \in G } \enc{n_g}) \parallel (\bigpar_{v_j \in V } \enc{v_j} ) \parallel (\bigpar_{n_c \in O } \enc{n_c})
		\]
		
	\paragraph{\bf Input generators.}
		The behaviour of input generator neurons is part of the specification of the network. Here we define two kinds of input behaviours: regular and non-deterministic ones. For each family, we  provide an encoding into timed automata.
		
		\paragraph{Regular input sequences.} Spike trains are ``regular" sequences of spikes and pauses: spikes are instantaneous while pauses have a non-null duration.
		Sequences can be \emph{empty}, \emph{finite} or \emph{infinite}.
		After each spike there must be a pause, except when the spike is the last event of a finite sequence.
		Infinite sequences are composed by two parts: a finite and arbitrary prefix and an infinite and periodic part composed by a finite sequence of \emph{spike--pause} pairs which is repeated infinitely often.
		More formally, such sequences are given in terms of the following grammar:
		$$
		\begin{array}{ll}
		G ::=& \Phi . (\Phi)^\omega \mid P(d) .  \Phi . (\Phi)^\omega \\
		\Phi ::= & s . P(d) . \Phi \mid \varepsilon
		\end{array}
		$$
		with $s$ representing a spike and $P(d)$ a pause of duration $d$. 
			It is possible to generate an emitter automaton for any regular input sequence:
	\begin{definition}[Input generator]
		Let $I \in \mathcal{L}(\nonterminal{G})$ be a word over the language generated by $IS$, then its encoding into timed automata is $\enc{I} = (L(I),\, first(I),\, \{ t \}$, $\{ y \},\, Arcs(I),\, Inv(I))$. It is inductively defined as follows:
		\begin{itemize}
			\item if $I := \Phi_1 . (\Phi_2)^\omega$ \begin{itemize}
				\item $L(I) = L(\Phi_1) \cup L(\Phi_2)$, where $last(\Phi_2)$ is \emph{urgent}
				\item $first(I) = first(\Phi_1)$
				\item $\arcs(I) = \arcs(\Phi_1) \cup \arcs(\Phi_2)\; \cup$\\
				$\{
					(last(\Phi_1), \true, \varepsilon, \emptyset, first(\Phi_2)),$\\
					$(last(\Phi_1), \true, \varepsilon, \emptyset, first(\Phi_2))
				\}$
				\item $\inv(I) = \inv(\Phi_1) \cup \inv(\Phi_2)$
			\end{itemize}
			
			\item if $I := P(d) . \Phi_1 . (\Phi_2)^\omega$ \begin{itemize}
				\item $L(I) = \{ \rstate{P}_0 \} \cup L(\Phi_1) \cup L(\Phi_2)$, where $last(\Phi_2)$ is \emph{urgent}
				\item $first(I) = \rstate{P}_0$
				\item $\arcs(I) = \arcs(\Phi_1) \cup \arcs(\Phi_2)\; \cup$\\
				$\{
				(\rstate{P}_0, t \leq d, ~, \{ t := 0 \}, first(\Phi_1))$,\\
				$(last(\Phi_1), \true, \varepsilon, \emptyset, first(\Phi_2)),$\\
				$(last(\Phi_1), \true, \varepsilon, \emptyset, first(\Phi_2))
				\}$
				\item $\inv(I) = \{ \rstate{P}_0 \mapsto t \leq d \} \cup \inv(\Phi_1) \cup \inv(\Phi_2)$
			\end{itemize}
			
			\item if $\Phi := \varepsilon$ \begin{itemize}
				\item $L(\Phi) = \{ \rstate{E} \}$
				\item $first(\Phi) = last(\Phi) = \rstate{E}$
				\item $\arcs(\Phi) = \emptyset$
				\item $\inv(\Phi) = \emptyset$
			\end{itemize}
			
			\item if $\Phi := s . P(d) . \Phi'$ \begin{itemize}
				\item $L(\Phi) = \{ \rstate{S}, \rstate{P} \} \cup L(\Phi')$
				\item $first(\Phi) = \rstate{S}, \quad last(\Phi) = last(\Phi')$
				\item $\arcs(\Phi) = \arcs(\Phi') \cup \{
					(\rstate{S}, \true, y!, \emptyset, \rstate{P})$,\\
					$(\rstate{P}, t = d, \varepsilon, \{ t := 0 \}, first(\Phi'))
				\}$
				\item $\inv(\Phi) = \{ \rstate{P} \mapsto t \leq d \} \cup \inv(\Phi')$
			\end{itemize}
		\end{itemize}
	\end{definition}
		
	\input{inputgen_encoding_intuition.tex}

	\paragraph{Non-deterministic input sequences.}		
		This kind of input sequences is useful when  no assumption is  available  on neuron inputs.
		These are random sequences of spikes separated by at least  $T_{min}$ time units.
		
		Such sequences can be generated by an automaton defined as follows:
		\begin{definition}[Non-deterministic input generator]
			A non-deterministic input generator $I_{nd}$ is a tuple $$(L,\rstate{B},X,\Sigma,\arcs,\inv),$$ with:
			\begin{itemize}
				\item $L = \{\rstate{B},\, \rstate{S},\, \rstate{W}\}$, with $\rstate{S}$ urgent,
				\item $X = \{t\}$
				\item $\Sigma = {x}$
				\item $Arcs = \{ (\rstate{B}, t = D, x!, \emptyset, \rstate{S}),\, (\rstate{S}, \true, \varepsilon, \{ t := 0 \}, \rstate{W}),\\ (\rstate{W}, t > T_{min}, x!, \emptyset, \rstate{S}) \}$
				\item $Inv(B) = (t \leq D)$
			\end{itemize}
			where $D$ is the initial delay.
		\end{definition}
		
		The behavior of such a generator depends on clock  $t$ and  broadcast channel $x$, and can be summarized as follows: it waits in location $\rstate{B}$ an arbitrary amount of time before moving to location $\rstate{S}$, firing its first spike over channel $x$.
		Since location \textbf{S} is \emph{urgent}, the automaton instantaneously moves to location $\rstate{W}$, resetting clock $t$.
		Finally, from location $\rstate{W}$, after an arbitrary amount of time $t$, 
		it moves to location $\rstate{S}$, firing a spike.
		Notice that an initial delay $D$ may be introduced by adding the invariant $t \leq D$ to the location \textbf{B} and the guard $t = D$ on the edge $\redge{B}{S}$.

%
		
\paragraph{\bf Standard neurons.}		
		The neuron is a computational unit behaving as follows:
			i) it accumulates potential whenever it receives input spikes within a given \emph{accumulation period},
			ii) if the accumulated potential is greater than the \emph{threshold}, it emits an output spike, iii) it waits during a  \emph{refractory period}, and restarts  from i).
			Observe that the accumulation period is not present in the definition of neuron (Definition \ref{def.snn}). It is indeed introduced here to  slice time and therefore discretise the  decrease of the potential value due to the leak factor.  
		We assume that two input spikes on the same synapse cannot be received within the same accumulation period (\ie the accumulation period is shorter than the minimum refractory period of the input neurons of the network). Next, we give the encoding of neurons into timed automata.
		
		\begin{definition}
			\label{def.nn.liaf.sync}
			Given a neuron $v=(\theta, \tau, \lambda, p, y)$ with $m$ input synapses, its encoding into timed automata is  ${\mathcal{N}} = (L,\rstate{A},X,Var,\Sigma,\arcs,\inv)$ with:
		
			\begin{itemize}
				\item $L = \{\rstate{A}, \rstate{W}, \rstate{D}\}$  with $\rstate{D}$ committed,
				\item $X=\{t\}$
				\item $Var=\{ p,a\}$
				\item $\Sigma = \{ x_i \mid i \in [1..m] \} \cup \{ y\}$,
				\item $\arcs=\{ (\rstate{A}, t\leq T, x_i?, \{a:=a+w_i\}, \rstate{A}) \mid i \in [1..m]\}
					\cup \{(\rstate{A}, t = T, ~, \{p:=a+\lfloor \lambda p \rfloor \}, \rstate{D}), \\
					(\rstate{D}, p < \theta, ~, \{a:=0\}, \rstate{A}),
					(\rstate{D}, p \geq \theta, y!, ~, \rstate{W}),\\
					(\rstate{W}, t=\tau , ~, \{a:=0, t:=0, p:=0\}, \rstate{A}) \}$  ;
			  \item $\inv(\rstate{A}) = t \leq T, \inv(\rstate{W}) = t \leq \tau, \inv(\rstate{D}) = \true$.
			\end{itemize}
			
		\end{definition}

\begin{figure}[t]
\centering
\newcommand{\sn}{0.8}	
	
\subfigure[	 Neuron model.]{\label{fig.ta.sync_neuron}

			\begin{tikzpicture}[scale = \sn]
			\node[draw, circle] (A) at (4,0) {A} ;
			\node[draw, circle, minimum width = 0.6cm](A2) at (4,0) {};
			\node[draw, circle] (D) at (4,-4) {D};
			\node[draw, circle] (W) at (0,-4) {W};
			
			\draw(4.5,-0.2) node[right,scale =\sn ] {\textbf{t $\leq$ T}};
			\draw(4.3,-4.2) node[right,scale =\sn ] {$C$}; 
			\draw(-0.5,-4.2) node[left]{\textbf{t $\leq\tau$}};

			\draw[->, >=latex] (A) -- (D);
			\draw (1, -1.5) node[right,scale =\sn ]{$G : t=T$} ;
			\draw (1, -1.9) node[right,scale =\sn ]{$S : -$} ;
			\draw (1, -2.3) node[right,scale =\sn ]{$U : p := a + \lfloor \lambda p \rfloor $};
			\draw (1.5, -2.7) node[right,scale =\sn ]{$t:= 0 $};
			
			\draw[->, >=latex] (D) -- (W); 	
			\draw (1.5, -4.2) node[right,scale =\sn ]{$G : p\geq\theta$} ;
			\draw (1.5, -4.6) node[right,scale =\sn ]{$S : y!$} ;
			\draw (1.5, -5) node[right,scale =\sn ]{$U : -$} ;

			\draw[->, >=latex] (W) to[bend left = 45]  (A); 
			\draw (-1, 0) node[right,scale =\sn ]{$G : t=\tau$} ;
			\draw (-1, -0.4) node[right,scale =\sn ]{$S : -$} ;
			\draw (-1, -0.8) node[right,scale =\sn ]{$U : t := 0$} ;
			\draw (-0.6, -1.2) node[right,scale =\sn ]{$a := 0$} ;
			\draw (-0.6, -1.6) node[right,scale =\sn ]{$p := 0$} ;

			\draw[->, >=latex] (A) to[out = 45, in = 0] (4,1) to[out = 180, in = 135 ] (A); 
			\draw (5, 1.8) node[right,scale = \sn ]{$\forall i = 1, ..., m$} ;
			\draw (5, 1.4 ) node[right,scale = \sn ]{$G: t \leq T$} ;
			\draw (5, 1) node[right, scale = \sn]{$S: x_i ?$};
			\draw (5, 0.6) node[right, scale = \sn]{$U: a:= a + w_i$};

			\draw[->, >=latex] (D) to[bend right = 45] (A); 
			\draw (5, -1.4) node[right,scale = \sn ]{$G : p<\theta$} ;
			\draw (5, -1.8) node[right,scale = \sn ]{$S : -$};
			\draw (5, -2.2) node[right,scale = \sn ]{$U : a = 0$};
		\end{tikzpicture} \linebreak

			}
\quad
\subfigure[Output consumer automaton.]{	\label{fig.ta.output_consumer}
			\begin{tikzpicture}[scale = \sn]
			\node[draw,circle] (W) at (0,0) {W} ;
			\node[draw, circle, minimum width = 0.6cm](W2) at (0,0) {};
			\node[draw, circle] (O) at (3,0){O};
			
			\draw(3.9, -0.3) node[left] {$U$};
			
			\draw[->, >=latex] (O) to[bend left = 45] (W); 
			\draw(1,1) node[right, scale = \sn]{$S : y?$};
			\draw[->, >=latex] (W) to[bend left = 45] (O);
			\draw(1,-1) node[right, scale = \sn]{$U : s := 0 $};
			\draw(1.5,-1.4) node[right, scale  = \sn]{$ e := not(e)$};
			
			\end{tikzpicture}
						}
\caption{Automata for standard neuron and output consumer.}
\label{fig.neuron-output_consumer}			
		\end{figure}

		The neuron behavior, described by the  automaton  in \figref{fig.ta.sync_neuron}, depends on the following channels, variables and clocks:
		\begin{itemize}
		 	\item $x_i$ for $i \in [1..m]$ are the $m$ input channels,
		 	\item $y$ is the broadcast channel used to emit the output spike,
		 	\item $p \in \mathbb{N}$ is the current potential value, initially set to zero,
	                \item $a \in \mathbb{N}$ is  the weighted sum of input spikes occurred within the \emph{current} accumulation period; it equals zero at the beginning of each round.
		\end{itemize}
		
	The behaviour  of the automaton modelling neuron $v$ can be summed up as follows:
		 \begin{itemize}
			\item the neuron keeps waiting in state \textbf{A} (for Accumulation) for input spikes while $t \leqslant T$ and, whenever it receives a spike on input $x_i$, it updates $a$ with $a := a + w_i$;
	
			\item when $t = T$, the neuron moves to state \textbf{D} (for Decision), resetting $t$ and updating $p$ according to the potential function given in Definition \ref{def.snn}:
			\[
				p := a + \left\lfloor \lambda \cdot p \right\rfloor
			\]
			Since state \textbf{D} is \emph{committed}, it does not allow time to progress, so, from this state, the neuron can move back to state \textbf{A} resetting $a$ if the potential has not reached the threshold $p < \theta$, or it can move to state \textbf{W}, firing an output spike, otherwise;
			\item the neuron remains in state \textbf{W} (for Wait) for $\tau$ time units {($\tau$ is the length of the refractory period)} and then it moves back to state \textbf{A} resetting $a$, $p$ and $t$.
		\end{itemize}
					
	\paragraph{\bf Output consumers.}

	In order to have a complete modelling of a \SNN, for each output neuron we build an \emph{output consumer} automaton $O_y$. The automaton,  whose formal definition is straightforward, is shown in \figref{fig.ta.output_consumer}. The consumer  waits in location \textbf{W} for the corresponding output spikes on channel $y$ and, as soon as it receives the spike, it moves to location \textbf{O}.
	This location is only needed to simplify model checking queries. Since it is urgent, the consumer instantly moves back to location \textbf{W} resetting $s$, the clock measuring the elapsed time since last emission, and setting $e$ to its negation, with  $e$ being  a \emph{boolean variable} which differentiates each emission from its successor.
	

\begin{definition}[Output consumer]
An output consumer is a timed automaton $${\mathcal{N}} = (L,\rstate{W},X,Var,\Sigma,\arcs,\inv)$$ with:

\begin{itemize}
				 \item $L = \{ \rstate{W}, \rstate{O}\}$ with O urgent, 

				\item $X=\{s\}$
				\item $Var=\{ e\}$
				\item $\Sigma = \{ y_i \mid y_i \text{ is an output neuron}\}$ 
				\item $\arcs=\{ 
				 (\rstate{W},  ,y?,  , \rstate{O}), \\
				 (\rstate{O}, s := 0, ,  \{e := not(e)\} ,\rstate{W}) \}$
				  \item $\inv(\rstate{W}) = \true, \inv(\rstate{O}) = \true$.
		
				\end{itemize}

\end{definition}

We have a complete implementation of the \SNN model proposed in the paper  via the tool \Uppaal. It can be found on the web page \cite{web}.
We have validated our neuron model against some characteristic properties studied in \cite{izhikevich04} (tonic spiking, excitability, integrator, etc.). These properties have been formalised  in temporal logics and checked via model-checking tools. 


Observe that, since we rely on a discrete time, we could have used tick automata \cite{tick}, a variant of B\"{u}chi automata where a special clock models the discrete flow of time. However, to the best of our knowledge, no existing tool allows to implement such automata. We decided to opt for timed automata in order to have an effective implementation of our networks to be exploited in parameter learning algorithms.

%% file: inputgen_encoding_intuition.tex

\begin{figure}
	\subfigure[$\enc{\Phi_1 . (\Phi_2)^\omega}$]{
		\label{fig.inputgen_encoding.root}
		\includegraphics[scale=0.3]{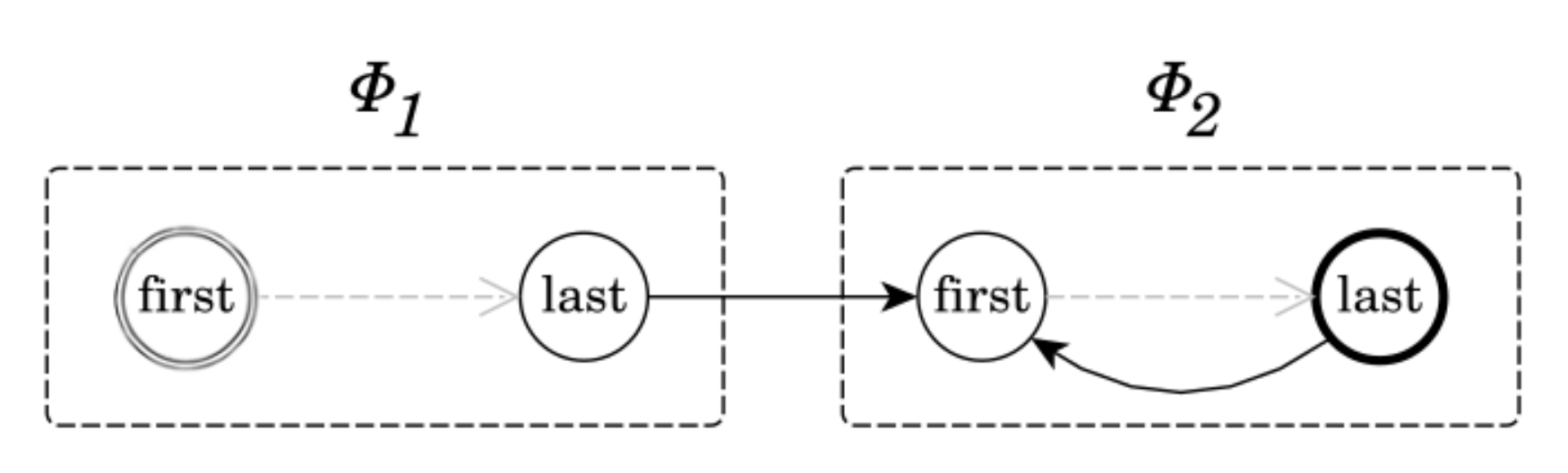}
		}
		\quad
	\subfigure[$\enc{P(d) . \Phi_1 . (\Phi_2)^\omega}$]{
		\label{fig.inputgen_encoding.pauseroot}
		\includegraphics[scale=0.3]{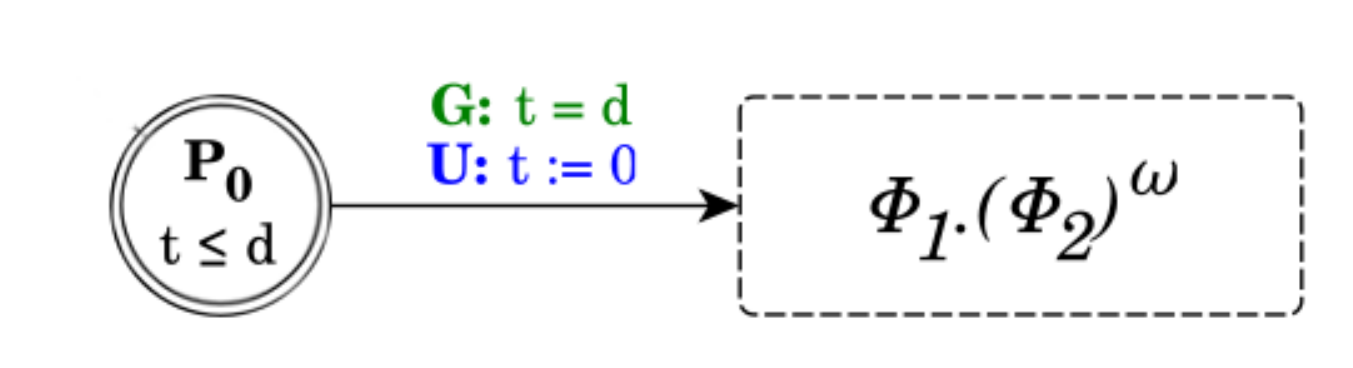}
		}
	
	\vspace{.5cm}
	
	\subfigure[$\enc{\varepsilon}$]{
		\label{fig.inputgen_encoding.empty}
		\includegraphics[scale=0.3]{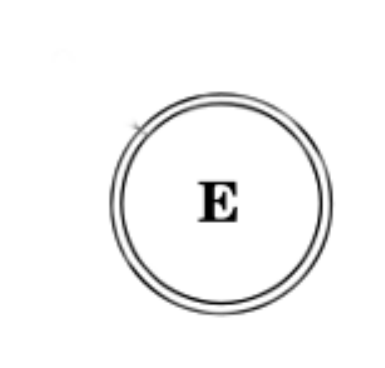}
}
	\subfigure[$\enc{s . P(d) . \Phi'}$]{
		\label{fig.inputgen_encoding.spikepause}
		\includegraphics[scale=0.3]{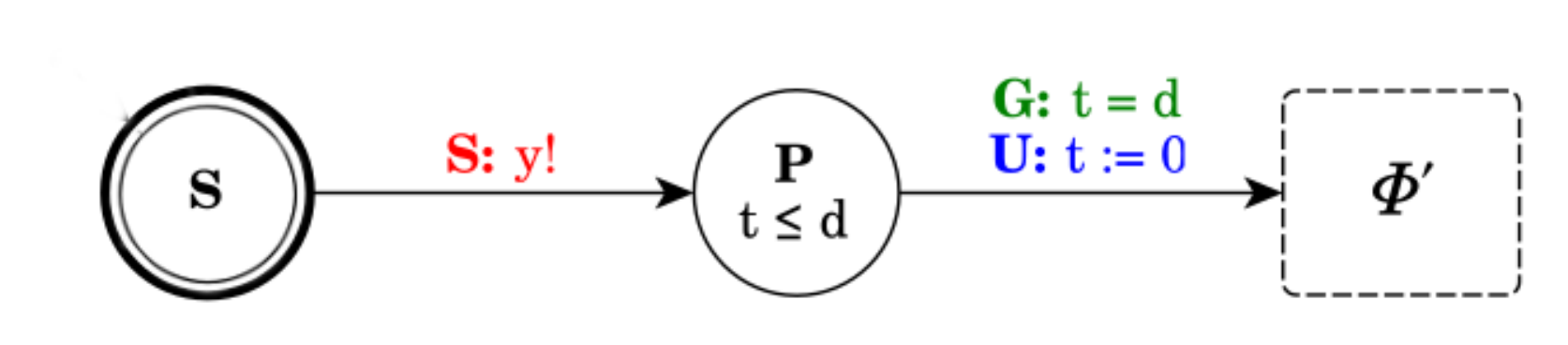}
		}

	\caption{Representation of the encoding of an input sequence}
	\label{fig.inputgen_encoding}
\end{figure}

\figref{fig.inputgen_encoding} depicts the shape of input generators.  Figure \ref{fig.inputgen_encoding.root} shows the generator $\enc{I}$, obtained from $I := \Phi_1 . (\Phi_2)^\omega$.
The edge connecting the last state of $\enc{\Phi_2}$ to the first one allows $\Phi_2$ to be repeated infinitely often.
Figure  \ref{fig.inputgen_encoding.pauseroot} shows the case of an input sequence $I := P(d) . \Phi_1 . (\Phi_2)^\omega$ beginning with a pause $P(d)$: in this case, the initial location of $\enc{I}$ is $\rstate{P}_0$, which imposes a delay of $d$ time units.
The remainder of the input sequence is encoded as for the previous case.
Figure \ref{fig.inputgen_encoding.empty} shows the induction basis for encoding a sequence $\Phi$, \ie the case $\Phi := \varepsilon$.
It is encoded as a location $\rstate{E}$ having no edge.
Finally, Figure \ref{fig.inputgen_encoding.spikepause} shows the case of a non-empty spike--pause pair sequence $\Phi := s.P(d).\Phi'$.  It consists of an \emph{urgent} location $\rstate{S}$: when the automaton moves from $\rstate{S}$, a spike is fired over channel $y$  and the automaton moves to location $\rstate{P}$, representing a silent period. After that, the automaton proceeds with the encoding of $\Phi'$.

%% file: inner.tex
In this section we show some properties of the neuron model of Definition \ref{def.nn.liaf.sync}. The first group of properties are structural. 
We  can compute a minimum  value  such that any neuron, having a threshold greater than or equal to it, will never be able to fire.

		\begin{property}
			\label{nn.props.theta_lambda_relation}
			Let $\ratuomaton{N} = (\theta, \tau, \lambda, p, y)$ be a neuron and $a_{max}$ be the maximum value received during each accumulation period. Then, if $\theta \geq \frac{a_{max}}{1 - \lambda}$, the neuron is not able to fire.
		\end{property}

\begin{proof}
			Without loss of generality, we suppose that, during each accumulation period, $\ratuomaton{N}$ receives the maximum possible input $a_{max}$. Then, its potential function is: \[
			p_n = a_{max} + \lfloor \lambda \cdot p_{n - 1} \rfloor
			\] which is always lower than or equal to its undiscretized version: \[
				p_n \leq {p'}_n = a_{max} + \lambda \cdot p'_{n - 1}
			\] The same inequality can be written in explicit form: \[
				p_n \leq {p'}_n = \sum_{k = 0}^{n} a_{n-k} \cdot \lambda^k
			\] and, since we assumed the neuron always receives $a_{max}$, $a_{n-k}$ is constant and does not depend on $k$: \[
				p_n \leq a_{max} \cdot \sum_{k = 0}^{n} \lambda^k
			\] The rightmost factor is a geometric series: \[
				p_n \leq a_{max} \cdot \frac{1 - \lambda^n}{1 - \lambda}
			\] which reaches its maximum value $\frac{1}{1 - \lambda}$ for $n \rightarrow \infty$, therefore: \[
				p_n \leq \frac{a_{max}}{1 - \lambda}.\ 
			\]
			Thus, if $\theta \geq \frac{a_{max}}{1 - \lambda}$, it is impossible for the neuron potential to reach the threshold and, consequently, the neuron cannot fire.
		\qed
		\end{proof}

In what follows, we only consider neurons that respect the previous constraint.		
			
		Apart from the minimum threshold, we can also quantify the amount of time that the neuron requires to complete an accumulate--fire--rest cycle. We show that there exists a minimum delay between neuron emissions.
		
		\begin{property}
			\label{nn.props.minimum_period}
			Let $\ratuomaton{N} = (\theta, \tau, \lambda, p, y)$ be a neuron. Then the time difference between successive firings cannot be lower than $T + \tau$.
		\end{property}
\begin{proof}
			Let $A_n = \sum_{k=1}^{T} a_{k + t_0}$ be the sum of weighted inputs during the $n$-th \emph{accumulation period}, then the neuron behaviour can be described as follows:
			\begin{equation*}
				\label{nn.props.eqs.pnAn}
				p_n = A_n + \lfloor \lambda \cdot p_{n - 1} \rfloor
			\end{equation*}
			which is the potential value after the $n$-th accumulation period. If the neuron eventually fires an output spike, then there exists $\hat{n} > 0$ such that:
			\begin{equation*} 
				\label{nn.props.eqs.n_star}
				\hat{n} = \underset{n \in \mathbb{N}}{arg\,min}\, \{ p_n : p_n \geq \theta \}
			\end{equation*}
			\ie{} the firing will occur at the end of the $\hat{n}$-th accumulation period, which means during the $\hat{t}$-th time unit since $t_0$, thus:
			\begin{equation*}
				\label{nn.props.eqs.t_star}
				\hat{t} = \hat{n} \cdot T + t_0
			\end{equation*}
			where $t_0$ is the \emph{last} reset time, \ie{} the last instant back in time when the neuron completed its refractory period.
			Then the \emph{next} reset time $t'$, \ie{} the next instant in future when the neuron will complete its refractory period, after having emitted a spike, is:
			\[
				t' = \hat{t} + \tau = \hat{n} \cdot T + \tau + t_0
			\]
			At instant $t'$, the neuron quits its refractory period, $n$ is reset to $0$, $t_0$ is set to $t'$, and $\hat{n}$, $\hat{t}$ and $t'$ must be consequently re-computed as described above.
			
			Such a way to describe our model dynamics allow us to express the \emph{inter-firing period} as a function of $\hat{n}$: \begin{equation*}
				\label{nn.props.eqs.ADW_loop_duration}
				t' - t_0 = \hat{n} \cdot T + \tau
			\end{equation*}
			So, the minimum inter-firing period is $T + \tau$ for $\hat{n} = 1$.\qed \end{proof}
			Such a property can also be verified as follows: let $\ratuomaton{I}$ be the non-deterministic input generator having $T_{min} = 1$ and, without loss of generality\footnote{the initial delay is required in order to make the formula hold for the first output spike too}, let the initial delay $D = T + \tau$. Then the \TAN{} $\ratuomaton{I} {\parallel} \ratuomaton{N}{\parallel} \ratuomaton{O}$ satisfies the following formula: \begin{equation*}
				AG (\pstate[O]{O} \rightarrow \pvar[O]{s} \geq T + \tau)
			\end{equation*} where $\rvar{s}$ measures the time elapsed since last firing, meaning that, whenever the output consumer receives a spike, the time elapsed since the previous received spike cannot be lower than $T + \tau$.

The next  fact states that only positive stimulations are necessary for the neuron to produce emissions.		
		\begin{fact}
			\label{nn.props.neg_inputs_are_inhibitory}
			Let $\ratuomaton{N} = (\theta, \tau, \lambda, p, y)$ be a  neuron,  $a(t)$  the sum of weighted inputs received during the current accumulation period, and $p(t-1)$  the neuron potential at the end of the previous accumulation period. If $p(t-1) < \theta$ and $a(t) < 0$, the neuron cannot fire at the end of the current accumulation period.
Moreover,  if $p(t) \geq \theta$ then $ a(t) > 0$.	
		\end{fact}

		The neuron potential is affected by every input spike it received \emph{since the last reset time}, but every event that occurred before that instant is forgotten: \ie neurons are memoryless.
		
		\begin{definition}[Inter-emission memory]
			Let $\ratuomaton{N}$ be a neuron,  $Z_\ratuomaton{N}$ its reset times set, and $I$ an input sequence.  Then $\ratuomaton{N}$ has inter-emission \emph{memory} if and only if there exist two different $t,t' \in Z_\ratuomaton{N}$ such that the output sequences produced by $\ratuomaton{N}$ as a response to $I$ starting from $t$ and $t'$ are different.
		\end{definition}
		

		\begin{property}
			\label{nn.props.no_memory}
			Neurons    
			have not inter-emission memory.
		\end{property}
		
		\begin{proof}
			When the neuron moves from location $\rstate{W}$ to $\rstate{A}$, it resets clock $t$ and variables $p$ and $a$, making them equal to their initial values. This entails that the neuron, if subjected to the same input sequence,  will always behave in the same way.
		\qed \end{proof}

%% file: n_props.tex

Next,  we validate the  neuron model against its ability of reproducing or not some behaviours, as described by Izhikevich in \cite{izhikevich04}. We introduce first three behaviours that are verified by our model.

	\paragraph{\bf Tonic Spiking.}	
		\emph{Tonic spiking} is the behaviour of a neuron producing a periodic output sequence as a response to a persistent excitatory constant input sequence.

%
%
%
%
%
%
		
		\begin{property}[Tonic spiking]
			\label{nn.props.tonic_spiking}	
			Let $\ratuomaton{N} =(\theta, \tau, \lambda, p, y)$ be a  neuron having only one ingoing excitatory synapse of weight  $w$ and let $\ratuomaton{I}$ be the input source connected to $\ratuomaton{N}$ producing a persistent input sequence. Then $\ratuomaton{N}$ produces a periodic output sequence.
		\end{property}
		

The property holds by construction. It can be tested via model checking in the following way.
			Let $\ratuomaton{I}$ be the fixed-rate input generator having arbitrary initial delay $D$, and let
			  $\ratuomaton{O}$ be an output consumer. Then the \TAN{} $\ratuomaton{I} {\parallel} \ratuomaton{N} {\parallel} \ratuomaton{O}$ satisfies the following formulae: \begin{equation*}				
				\begin{cases}
					 \pstate[O]{O} \wedge \pvar[O]{e} \leadsto \pstate[O]{O} \wedge \neg \pvar[O]{e} \\
					\pstate[O]{O} \wedge \neg \pvar[O]{e} \leadsto \pstate[O]{O} \wedge \pvar[O]{e}
				\end{cases}
			\end{equation*} where $\rstate{O}$ is the location that the consumer automaton $\ratuomaton{O}$ reaches after consuming a spike and $\rvar{e}$ is an alternating boolean variable whose value flips whenever $\ratuomaton{O}$ moves into location $\rstate{O}$.
			So, whenever automaton $\ratuomaton{O}$ reaches location $\rstate{O}$, it will eventually reach it again.
			
	One may also find the value $P$ of the period of some given neuron $\ratuomaton{N}$ by means of simulations, thus the periodic behaviour can be proven verifying the following formula: \begin{equation*}
			AG ( \pstate[O]{O} \wedge \pvar[N]{f} \rightarrow \pvar[O]{s} = P )
		\end{equation*} where $\rvar{s}$ is the clock measuring the time elapsed since last spike consumed by $\ratuomaton{O}$, and $\rvar{f}$ is a boolean variable of automaton $\ratuomaton{N}$ which is initially $false$ and is set to $true$ when edge $\redge{W}{A}$ fires (\ie{} it indicates whether $\ratuomaton{N}$ has already emitted the first spike and waited the first refractory period or not).
		
	\paragraph{\bf Integrator.}
		\emph{Integrator} is the behaviour of a neuron producing an output spike whenever it receives \emph{at least} a specific number of  spikes from its input sources in the same accumulation period. 		
		\begin{property}[Integrator]		
			Let $\ratuomaton{N} = (\theta, \tau, \lambda, p, y)$ be a  neuron having $m$ synapses with maximum excitatory weight $R$ and a threshold $n \leq m$. Then the neuron emits if it receives a spike from at least $n$ input sources during the same accumulation period.
		\end{property}
		
%
%
%
%
		

As in the previous case, we can use model checking tools and test the formula  stating that, if at least $n$ generators are ready to emit (location $\rstate{S}$) while $\ratuomaton{N}$ is in $\rstate{A}$, then $\ratuomaton{O}$ will eventually capture an output of $\ratuomaton{N}$: \begin{equation*}
				\left( \sum_{i=1}^{m} \pstate[][i]{S} \geq n \right) \wedge \pstate[\ratuomaton{N}]{A} \leadsto \pstate[\ratuomaton{O}]{O}
			\end{equation*}
					
		Notice that, since potential depends on past inputs too, the neuron may still be able to fire in other circumstances, \eg if it keeps receiving less than $n$ spikes for a sufficient number of accumulation periods, then it may eventually fire.
		
%
%
		
	\paragraph{\bf Excitability.}
		\emph{Excitability} is the behaviour of a neuron emitting sequences having a \emph{decreasing} inter-firing period, \ie an increasing output frequency, when stimulated by an \emph{increasing} number of excitatory inputs. 		
		\begin{property}[Excitability]
			Let $\ratuomaton{N} = (\theta, \tau, \lambda, p, y)$ be a neuron having $m$ excitatory synapses. Then the inter-spike period decreases as the sum of weighted input spikes increases.
		\end{property}
		
		\begin{proof}
			If we assume the neuron is receiving an increasing number of excitatory spikes, generated by an increasing number of input sources emitting persistent inputs, then $a_t$ is the non-negative, non-decreasing and progressing (\ie $\forall u\  \exists t : a_t > u$) succession representing the weighted sum of inputs within the $t$-th time unit. \
			Consequently, $A_n = \sum_{k=1}^{T} a_{k + t_0}$ is the non-negative, non-decreasing and progressing succession counting the total sum of inputs within the $n$-th accumulation period.
			Since $A_n$ is positive and  Property \ref{nn.props.minimum_period} holds, we can prove that the inter-spike period  $t_n - t_{n-1}$ decreases.
		\qed \end{proof}
	
The following behaviours are not satisfied by the \lif model, we show that our encoding cannot verify them as well.

	\paragraph{\bf Phasic Spiking.}
		\emph{Phasic spiking} is the behaviour of a neuron producing a \emph{single} output spike when receiving  a persistent and excitatory input sequence and then remaining quiescent for the rest of it.
Such a behaviour depends on the neuron to have inter-emission memory.

		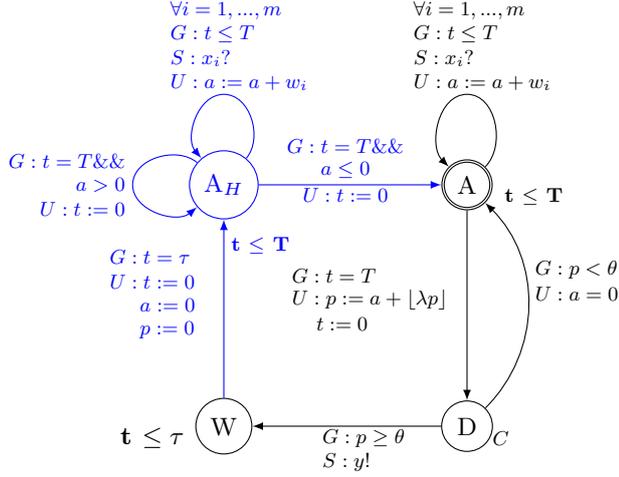
\begin{figure}
			\centering
			
			\newcommand{\sn}{0.8}

		\begin{tikzpicture}[scale = \sn]
			\node[draw, circle] (A) at (4,0) {A} ;
			\node[draw, circle, minimum width = 0.6cm](A2) at (4,0) {};
			\node[draw, circle] (D) at (4,-4) {D};
			\node[draw, circle] (W) at (0,-4) {W};
			\node[blue, draw, circle] (AH) at (0,0) {A$_H$};
			
			\draw(4.5,-0.2) node[right,scale =\sn ] {\textbf{t $\leq$ T}};
			\draw[blue](0,-1) node[right,scale =\sn ] {\textbf{t $\leq$ T}};
			\draw(4.3,-4.2) node[right,scale =\sn ] {$C$}; 
			\draw(-0.5,-4.2) node[left]{\textbf{t $\leq\tau$}};

			\draw[->, >=latex] (A) -- (D);
			\draw (1, -1.5) node[right,scale =\sn ]{$G : t=T$} ;
			\draw (1, -1.9) node[right,scale =\sn ]{$U : p := a + \lfloor \lambda p \rfloor $};
			\draw (1.4, -2.3) node[right,scale =\sn ]{$t:= 0 $};
			
			\draw[->, >=latex] (D) -- (W); 	
			\draw (1.5, -4.2) node[right,scale =\sn ]{$G : p\geq\theta$} ;
			\draw (1.5, -4.6) node[right,scale =\sn ]{$S : y!$} ;

			\draw[blue,->, >=latex] (W) --  (AH); 
			\draw[blue] (-2, -1.2) node[right,scale =\sn ]{$G : t=\tau$} ;
			\draw[blue] (-2, -1.6) node[right,scale =\sn ]{$U : t := 0$} ;
			\draw[blue] (-1.5, -2) node[right,scale =\sn ]{$a := 0$} ;
			\draw[blue] (-1.5, -2.4) node[right,scale =\sn ]{$p := 0$} ;	

			\draw[blue, ->, >=latex] (AH) -- (A); 
			\draw[blue] (2, 0.4) node[above, scale = \sn] {$G : t = T \&\&$}; 
			\draw[blue] (2, 0) node[above, scale = \sn] {$a \leq 0$};
			\draw[blue] (2, -0.4) node[above, scale = \sn] {$U : t := 0 $};

			\draw[->, >=latex] (A) to[out = 45, in = 0] (4,1.5) to[out = 180, in = 135 ] (A); 
			\draw (3, 2.9) node[right,scale = \sn ]{$\forall i = 1, ..., m$} ;
			\draw (3, 2.5 ) node[right,scale = \sn ]{$G: t \leq T$} ;
			\draw (3, 2.1) node[right, scale = \sn]{$S: x_i ?$};
			\draw (3, 1.7) node[right, scale = \sn]{$U: a:= a + w_i$};

			\draw[->, >=latex] (D) to[bend right = 45] (A); 
			\draw (5, -1.4) node[right,scale = \sn ]{$G : p<\theta$} ;
			\draw (5, -1.8) node[right,scale = \sn ]{$U : a = 0$};
			
			\draw[blue, ->, >=latex] (AH) to[out = 50, in = 0] (0,1.5) to[out = 180, in = 130 ] (AH); 
			\draw[blue] (-1, 2.9) node[right,scale = \sn ]{$\forall i = 1, ..., m$} ;
			\draw[blue] (-1, 2.5 ) node[right,scale = \sn ]{$G: t \leq T$} ;
			\draw[blue] (-1, 2.1) node[right, scale = \sn]{$S: x_i ?$};
			\draw[blue] (-1, 1.7) node[right, scale = \sn]{$U: a:= a + w_i$};

			\draw[blue, ->, >=latex] (AH) to[out = 140, in = 90] (-1.5,0) to[out = 270, in = 220 ] (AH); 
			
			\draw[blue] (-1.5, 0.4) node[left, scale = \sn] {$G : t = T \&\&$}; 
			\draw[blue] (-1.5, 0) node[left, scale = \sn] {$a > 0$};
			\draw[blue] (-1.5, -0.4) node[left, scale = \sn] {$U : t := 0 $};
			
		\end{tikzpicture} \linebreak

				\caption{
					The  extended neuron model for  phasic spiking.\\
					Additions are colored in blue.
				}
				\label{fig.automata.phasic_spiking}
		\end{figure}

		\begin{property}
			Neurons cannot reproduce the \emph{phasic spiking} behaviour.
		\end{property}
		
		\begin{proof}
			The phasic spiking behaviour requires the neuron to ignore any excitatory input spike occurring after its first emission. This means producing different outcomes, before and after the first emission, as a response to the same input sequence, which is impossible for a memoryless neuron, as stated in \propref{nn.props.no_memory}.
		\qed \end{proof}
	
	We can extend our model  to reproduce phasic spiking, see \figref{fig.automata.phasic_spiking}.
		This variant  makes the neuron able to ``remember'' if it is receiving a persistent excitatory input sequence.
		After each refractory period, the neuron moves to location $\rstate{A_H}$, instead of  $\rstate{A}$.
		The only difference between $\rstate{A_H}$ and  $\rstate{A}$ is that $\rstate{A_H}$ \emph{ignores} positive values of $a$ at the end of each accumulation period.
		Conversely, a non-positive value of $a$ (denoting the end of the persistent input), at the end of some accumulation period, leads the neuron back in location $\rstate{A}$.
				
	\paragraph{\bf  Bursting.}
		\label{par.tonic_bursting}
		A \emph{burst} is a finite sequence of \emph{high frequency} spikes. More formally:
		
		\begin{definition}
			A spike output sequence is a \emph{burst} if it is composed by  spikes having an occurrence rate greater than $1/\tau$, with $\tau$ being the refractory period of the neuron.
	
			A \emph{burst sequence} is a  sequence composed by bursts, subject to the following constraint: the time difference between the last spike of each burst and the first spike of the next burst it greater than $\tau$.
		\end{definition}
		
		\begin{property}
			\label{nn.props.no_bursts}
			Neurons  cannot produce bursts.
		\end{property}
		
		\begin{proof}
			A neuron $\ratuomaton{N}$ cannot emit spikes having a rate greater than $1/(T + \tau)$, as stated by \propref{nn.props.minimum_period}, so it cannot produce bursts.
		\qed \end{proof}

	In order to reproduce bursts our model can be extended by allowing several subsequent emissions in an interval period smaller than $\tau$. After this period all clocks and variables are reset and the accumulation-fire-rest cycle can start again.	
		
Several bursting behaviours are described in \cite{izhikevich04}. Here we discuss only three of them, as all impossibility results depend on Property \ref{nn.props.no_bursts} and all the automata extensions are similar.
		
		 \emph{Tonic Bursting} is the behaviour of a neuron producing a burst sequence as a response to a persistent and excitatory input sequence.
		\emph{Phasic Bursting} is the behaviour of a neuron producing a burst as a consequence  of a persistent excitatory input sequence and then remaining quiescent. Obviously the preceding behaviours require the ability of producing bursts.

		\emph{Bursting-then-Spiking} is the behaviour of a neuron producing a burst as response to a persistent excitatory input sequence and then producing a periodic output sequence.
		Such a behaviour, similarly to Phasic and Tonic Bursting, depends on the neuron ability of  producing bursts. Moreover it requires inter-emission memory, in order to detect the beginning of a persistent sequence.

		\begin{property}
			Neurons cannot exhibit the \emph{Tonic Bursting, Phasic Bursting} and \emph{Bursting-then-Spiking} behaviours.
		\end{property}
		
		\begin{proof}
			Follows from  \propref{nn.props.no_bursts}.
		\qed \end{proof}

	\paragraph{\bf Spike Frequency Adaptation.}
		\emph{Spike Frequency Adaptation} is the behaviour of a neuron producing a decreasing-frequency output sequence as a response to a persistent excitatory input sequence. In other words, the inter-emission time difference increases as the time elapses.
		This behaviour requires the neuron to have inter-emission memory as it should be able to keep track of the time elapsed since the beginning of the input sequence.
		
		\begin{property}
			Neurons cannot reproduce the \emph{Spike Frequency Adaptation} behaviour.
		\end{property}		
		\begin{proof}
			The Spike Frequency Adaptation behaviour requires the neuron to detect the beginning  of an excitatory input sequence and to increase the time required to fire a spike, after each emission.
			This means the neuron will produce different outcomes as response to equal inputs, which is impossible, as stated in \propref{nn.props.no_memory}.
		\qed \end{proof}

		\begin{figure}
							\centering

			\newcommand{\sn}{0.8}	
	
			\begin{tikzpicture}[scale = \sn]
			\node[draw, circle] (A) at (4,0) {A} ;
			\node[draw, circle, minimum width = 0.6cm](A2) at (4,0) {};
			\node[draw, circle] (D) at (4,-4) {D};
			\node[draw, circle] (W) at (0,-4) {W};
			
			\draw(4.5,-0.2) node[right,scale =\sn ] {\textbf{t $\leq$ T}};
			\draw(4.3,-4.2) node[right,scale =\sn ] {$C$}; 
			\draw(-0.5,-4.2) node[left]{\textbf{t $\leq\tau$}};

			\draw[->, >=latex] (A) -- (D);
			\draw (1, -1.9) node[right,scale =\sn ]{$G : t=T$} ;
			\draw (1, -2.3) node[right,scale =\sn ]{$U : p := a + \lfloor \lambda p \rfloor $};
			\draw (1.4, -2.7) node[right,scale =\sn ]{$t:= 0 $};
			
			\draw[->, >=latex] (D) -- (W); 	
			\draw (1.5, -4.2) node[right,scale =\sn ]{$G : p\geq\theta$} ;
			\draw (1.5, -4.6) node[right,scale =\sn ]{$S : y!$} ;

			\draw[->, >=latex] (W) to[bend left = 45]  (A); 
			\draw (-0.5, 0.4) node[right,scale =\sn ]{$G : t=\tau$} ;
			\draw (-0.5, 0) node[right,scale =\sn ]{$U : t := 0$} ;
			\draw (0.1, -0.4) node[right,scale =\sn ]{$a := 0$} ;
			\draw (0.1, -0.8) node[right,scale =\sn ]{$p := 0$} ;	
			\draw[blue] (0.1, -1.2) node[right,scale =\sn ]{$\tau := \tau + \Delta \tau$} ;

			\draw[->, >=latex] (A) to[out = 45, in = 0] (4,1.5) to[out = 180, in = 135 ] (A); 
			\draw (5, 1.8) node[right,scale = \sn ]{$\forall i = 1, ..., m$} ;
			\draw (5, 1.4 ) node[right,scale = \sn ]{$G: t \leq T$} ;
			\draw (5, 1) node[right, scale = \sn]{$S: x_i ?$};
			\draw (5, 0.6) node[right, scale = \sn]{$U: a:= a + w_i$};

			\draw[->, >=latex] (D) to[bend right = 45] (A); 
			\draw (5, -1.8) node[right,scale = \sn ]{$G : p<\theta$} ;
			\draw (5, -2.2) node[right,scale = \sn ]{$U : a = 0$};
			\draw[blue](5.5, -2.6) node[right, scale = \sn]{$\tau := \tau_0$};
		\end{tikzpicture} \linebreak

				\caption{
					The extended  model for Spike Frequency Adaptation behaviour.
					Additions are colored in blue.
				}
				\label{fig.automata.spike_frequency_adaptation}
					\end{figure}

		An extended neuron model able to reproduce Spike Frequency Adaptation behaviour is  shown in \figref{fig.automata.spike_frequency_adaptation}.
		This variant allows the refractory period to increase after each neuron emission, thus making the output frequency decrease.
		
	\paragraph{\bf Spike Latency.}
		\emph{Spike Latency} is the behaviour of a neuron firing delayed spikes, with respect to the instant when its potential reached or overcame the threshold.
		Such a delay is proportional to the strength of the signal which leads it to emission, \ie the sum of weighed inputs received during the accumulation period preceding the emission. 		
		This behaviour requires the neuron to be able to postpone its output.
		
		\begin{property}
			Neurons cannot reproduce the \emph{Spike Latency} behaviour.
		\end{property}
		\begin{proof}
			The property holds by construction. As location $\rstate{D}$ is \emph{committed}, no firing can be delayed.
					\qed \end{proof}
		
		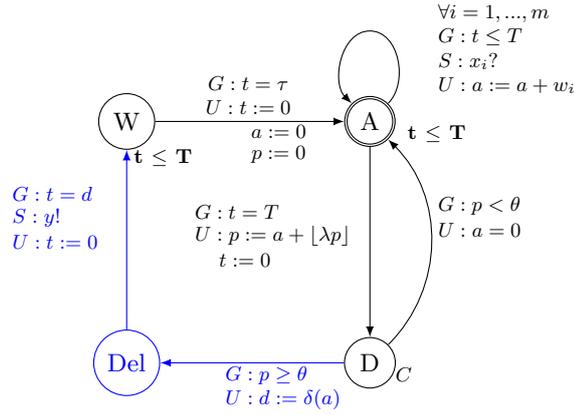
\begin{figure}
	
				\centering

				\newcommand{\sn}{0.8}	
				
		\begin{tikzpicture}[scale = \sn]
			\node[draw, circle] (A) at (4,0) {A} ;
			\node[draw, circle, minimum width = 0.6cm](A2) at (4,0) {};
			\node[draw, circle] (D) at (4,-4) {D};
			\node[blue, draw, circle] (Del) at (0,-4) {Del};
			\node[draw, circle] (W) at (0,0) {W};
			
			\draw(4.5,-0.2) node[right,scale =\sn ] {\textbf{t $\leq$ T}};
			\draw(0,-0.6) node[right,scale =\sn ] {\textbf{t $\leq$ T}};
			\draw(4.3,-4.2) node[right,scale =\sn ] {$C$}; 

			\draw[->, >=latex] (A) -- (D);
			\draw (1, -1.5) node[right,scale =\sn ]{$G : t=T$} ;
			\draw (1, -1.9) node[right,scale =\sn ]{$U : p := a + \lfloor \lambda p \rfloor $};
			\draw (1.4, -2.3) node[right,scale =\sn ]{$t:= 0 $};
			
			\draw[blue, ->, >=latex] (D) -- (Del); 	
			\draw[blue] (1.5, -4.2) node[right,scale =\sn ]{$G : p\geq\theta$} ;
			\draw[blue] (1.5, -4.6) node[right,scale =\sn ]{$U : d := \delta (a)$} ;

			\draw[blue, ->, >=latex] (Del) --  (W); 
			\draw[blue](-2, -1.2) node[right,scale =\sn ]{$G : t = d$} ;
			\draw[blue](-2, -1.6) node[right,scale =\sn ]{$S : y!$} ;
			\draw[blue] (-2, -2) node[right,scale =\sn ]{$U : t := 0$} ;	

			\draw[ ->, >=latex] (W) -- (A); 
			\draw (2, 0.4) node[above, scale = \sn] {$G : t = \tau$}; 
			\draw(2, 0) node[above, scale = \sn] {$U : t := 0$};
			\draw (2.5, -0.4) node[above, scale = \sn] {$ a := 0 $};
			\draw (2.5, -0.8) node[above, scale = \sn] {$ p := 0 $};

			\draw[->, >=latex] (A) to[out = 45, in = 0] (4,1.5) to[out = 180, in = 135 ] (A); 
			\draw (5, 1.8) node[right,scale = \sn ]{$\forall i = 1, ..., m$} ;
			\draw (5, 1.4 ) node[right,scale = \sn ]{$G: t \leq T$} ;
			\draw (5, 1) node[right, scale = \sn]{$S: x_i ?$};
			\draw (5, 0.6) node[right, scale = \sn]{$U: a:= a + w_i$};

			\draw[->, >=latex] (D) to[bend right = 45] (A); 
			\draw (5, -1.4) node[right,scale = \sn ]{$G : p<\theta$} ;
			\draw (5, -1.8) node[right,scale = \sn ]{$U : a = 0$};

		\end{tikzpicture} \linebreak

				\caption{
					The extended  model for the Spike Latency behaviour.
					Additions are colored in blue.
				}
				\label{fig.automata.spike_latency}			
		\end{figure}
		
An easy solution to extend our model is to  introduce a delay between the instant the neuron reaches or overcomes its threshold and the actual emission instant.
		Such a delay $\delta$ depends  on the sum of weighted inputs received during  the last accumulation period.
			If the potential is greater than or equal to the threshold, the neuron computes the delay duration $\delta(a)$, assigning it to an integer variable $d$, and then waits in location $\rstate{Del}$ for $d$ time units before emitting a spike on channel $y$.
			The extended version is depicted in 	 \figref{fig.automata.spike_latency}.
 		
	\paragraph{\bf Threshold Variability.}
		\emph{Threshold variability} is the behaviour of a neuron allowing its threshold to vary according to the strength of its inputs.
		More precisely, an excitatory input will rise the threshold while an inhibitory input will decrease it.
		As a consequence, excitatory inputs may more easily lead the neuron to fire when occurring after an inhibitory input.

		\begin{property}
			Neurons cannot reproduce the \emph{Threshold Variability} behaviour.
		\end{property}
		
		\begin{proof}
			By construction the neuron threshold never changes.
		\qed \end{proof}
	
	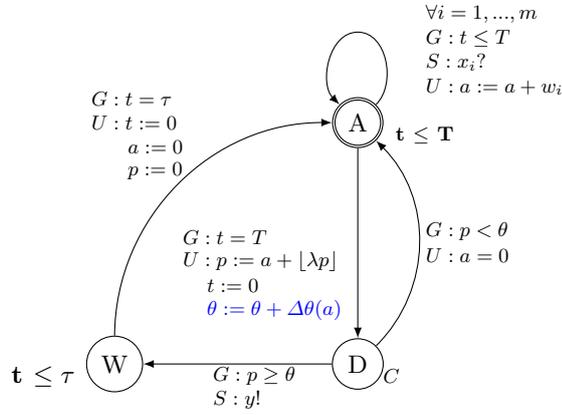
\begin{figure}
			\centering

				\newcommand{\sn}{0.8}	
	
			\begin{tikzpicture}[scale = \sn]
			\node[draw, circle] (A) at (4,0) {A} ;
			\node[draw, circle, minimum width = 0.6cm](A2) at (4,0) {};
			\node[draw, circle] (D) at (4,-4) {D};
			\node[draw, circle] (W) at (0,-4) {W};
			
			\draw(4.5,-0.2) node[right,scale =\sn ] {\textbf{t $\leq$ T}};
			\draw(4.3,-4.2) node[right,scale =\sn ] {$C$}; 
			\draw(-0.5,-4.2) node[left]{\textbf{t $\leq\tau$}};

			\draw[->, >=latex] (A) -- (D);
			\draw (1, -1.9) node[right,scale =\sn ]{$G : t=T$} ;
			\draw (1, -2.3) node[right,scale =\sn ]{$U : p := a + \lfloor \lambda p \rfloor $};
			\draw (1.4, -2.7) node[right,scale =\sn ]{$t:= 0 $};
			\draw[blue] (1.4, -3.1) node[right,scale =\sn ]{$\theta := \theta + \Delta\theta(a) $};

			\draw[->, >=latex] (D) -- (W); 	
			\draw (1.5, -4.2) node[right,scale =\sn ]{$G : p\geq\theta$} ;
			\draw (1.5, -4.6) node[right,scale =\sn ]{$S : y!$} ;

			\draw[->, >=latex] (W) to[bend left = 45]  (A); 
			\draw (-0.5, 0.4) node[right,scale =\sn ]{$G : t=\tau$} ;
			\draw (-0.5, 0) node[right,scale =\sn ]{$U : t := 0$} ;
			\draw (0.1, -0.4) node[right,scale =\sn ]{$a := 0$} ;
			\draw (0.1, -0.8) node[right,scale =\sn ]{$p := 0$} ;

			\draw[->, >=latex] (A) to[out = 45, in = 0] (4,1.5) to[out = 180, in = 135 ] (A); 
			\draw (5, 1.8) node[right,scale = \sn ]{$\forall i = 1, ..., m$} ;
			\draw (5, 1.4 ) node[right,scale = \sn ]{$G: t \leq T$} ;
			\draw (5, 1) node[right, scale = \sn]{$S: x_i ?$};
			\draw (5, 0.6) node[right, scale = \sn]{$U: a:= a + w_i$};

			\draw[->, >=latex] (D) to[bend right = 45] (A); 
			\draw (5, -1.8) node[right,scale = \sn ]{$G : p<\theta$} ;
			\draw (5, -2.2) node[right,scale = \sn ]{$U : a = 0$};
		\end{tikzpicture} \linebreak

				\caption{
					The extended  model for  the Threshold variability behaviour.
					Additions colored are in blue.				}
				\label{fig.automata.threshold_variability}
			
		\end{figure}
		
		The neuron model can be extended allowing the threshold to vary after each accumulation period according to the current sum of weighted inputs (see  \figref{fig.automata.threshold_variability}).
		The threshold variable initial value is $\theta_0$.
		On every firing of edge $\redge{A}{D}$, the threshold variable is increased of $\Delta(a)$, where $a$ is the sum of weighted inputs occurred during the last accumulation period and $\Delta(a)$  is an integer value whose sign is opposite to the sign of $a$ and whose magnitude is proportional to the magnitude of $a$.
		
	\paragraph{\bf Bistability.}
		\emph{Bistability} is the behaviour of a neuron alternating between two operation modes: \emph{periodic emission} and \emph{quiescence}. Upon reception of a single excitatory spike,  it emits a periodic output sequence and switches to a  quiescent mode (no emission) as soon as it received another spike.
		Such a behaviour requires the neuron to \begin{enumerate*}[label=(\roman*)]
			\item be able to produce a periodic output sequence, even if no excitatory spike is received,
			\item be able to remain silent when no spike is received, and
			\item be able to switch between the two operation modes upon reception of  an excitatory spike.
		\end{enumerate*}
		
		\begin{property}
			Neurons cannot reproduce the \emph{Bistability} behaviour.
		\end{property}
		
		\begin{proof}
	 The only possibility of obtaining a periodic output as a result of no excitatory input spike is to set  $\theta = 0$. This  is a limit case of \propref{nn.props.tonic_spiking}.
			Since, by construction, the threshold cannot vary,  the neuron cannot switch between the two operation modes.
		\qed \end{proof}

		\begin{figure}
			\centering

				\newcommand{\sn}{0.8}	
	
			\begin{tikzpicture}[scale = \sn]
			\node[draw, circle] (A) at (4,0) {A} ;
			\node[draw, circle, minimum width = 0.6cm](A2) at (4,0) {};
			\node[draw, circle] (D) at (4,-4) {D};
			\node[draw, circle] (W) at (0,-4) {W};
			
			\draw(4.5,-0.2) node[right,scale =\sn ] {\textbf{t $\leq$ T}};
			\draw(4.3,-4.2) node[right,scale =\sn ] {$C$}; 
			\draw(-0.5,-4.2) node[left]{\textbf{t $\leq\tau$}};

			\draw[->, >=latex] (A) -- (D);
			\draw (1, -1.9) node[right,scale =\sn ]{$G : t=T$} ;
			\draw (1, -2.3) node[right,scale =\sn ]{$U : p := a + \lfloor \lambda p \rfloor $};
			\draw (1.5, -2.7) node[right,scale =\sn ]{$t:= 0 $};
			\draw[blue] (1.5, -3.1) node[right,scale =\sn ]{$\theta := bist(\theta, a) $};

			\draw[->, >=latex] (D) -- (W); 	
			\draw (1.5, -4.2) node[right,scale =\sn ]{$G : p\geq\theta$} ;
			\draw (1.5, -4.6) node[right,scale =\sn ]{$S : y!$} ;

			\draw[->, >=latex] (W) to[bend left = 45]  (A); 
			\draw (-0.5, 0.4) node[right,scale =\sn ]{$G : t=\tau$} ;
			\draw (-0.5, 0) node[right,scale =\sn ]{$U : t := 0$} ;
			\draw (0.1, -0.4) node[right,scale =\sn ]{$a := 0$} ;
			\draw (0.1, -0.8) node[right,scale =\sn ]{$p := 0$} ;

			\draw[->, >=latex] (A) to[out = 45, in = 0] (4,1.5) to[out = 180, in = 135 ] (A); 
			\draw (5, 1.8) node[right,scale = \sn ]{$\forall i = 1, ..., m$} ;
			\draw (5, 1.4 ) node[right,scale = \sn ]{$G: t \leq T$} ;
			\draw (5, 1) node[right, scale = \sn]{$S: x_i ?$};
			\draw (5, 0.6) node[right, scale = \sn]{$U: a:= a + w_i$};

			\draw[->, >=latex] (D) to[bend right = 45] (A); 
			\draw (5, -1.8) node[right,scale = \sn ]{$G : p<\theta$} ;
			\draw (5, -2.2) node[right,scale = \sn ]{$U : a = 0$};
		\end{tikzpicture} \linebreak

				\caption{
					The extended  model for Bistability behaviour.\\
					Additions are colored in blue.
				}
				\label{fig.automata.bistability}
			
		\end{figure}
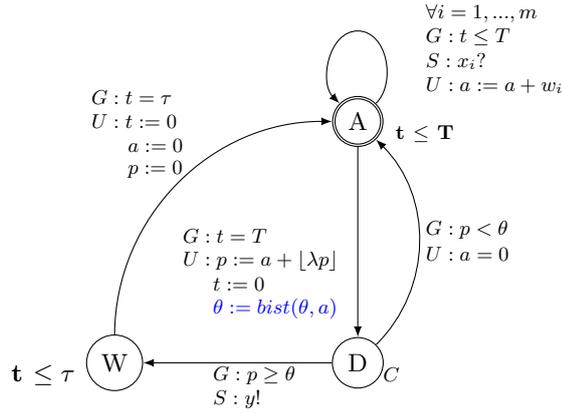

		The neuron model can be modified as shown in \figref{fig.automata.bistability}.
		This variant  makes its threshold switch between $0$ and a positive value at the end of any accumulation period during which it received an excitatory sum of weighted inputs $a$.
		A null threshold would make the neuron emit even if no input is received.
		Conversely, a positive threshold would prevent the neuron from emitting, if no input is received.
		Thus, on every firing of edge $\redge{A}{D}$, the threshold value $\theta$ is computed by  the function $bist(\cdot)$: \[
			bist(\theta,\, a) = \begin{cases}
				0 & if\ \theta > 0 \wedge a > 0 \\
				\theta_0 & if\ \theta = 0 \wedge a > 0 \\
				\theta & if\ a \leq 0.
			\end{cases}
		\]
		
	\paragraph{\bf Inhibition-induced activities.}
		This  is the behaviour of a neuron producing a spike  output sequence as a response to a persistent \emph{inhibitory} input sequence.
		We thus require the neuron to be able to emit as a consequence of some inhibitory input spikes. 		
		\begin{property}
			Neurons cannot reproduce the \emph{Inhibition-induced Spiking} behavior.
		\end{property}
		
		\begin{proof}
			Follows from Fact \ref{nn.props.neg_inputs_are_inhibitory}.
		\qed \end{proof}
		
		An easy extension to our automata is to consider  \emph{the absolute value} of all inputs instead of their signed values.
				
	\paragraph{\bf Rebound activities.}
		
		\emph{Rebound Spike}  is the behaviour of a neuron producing an output spike  after it received an inhibitory input.			
		Similarly to Inhibition-induced activities, this behaviour requires the neuron  to emit as a consequence of an inhibitory input spike.
		
		\begin{property}
			Neurons cannot exhibit the \emph{Rebound Spiking} behaviour.		\end{property}
		
			\begin{proof}
			Follows from Fact \ref{nn.props.neg_inputs_are_inhibitory}.
		\qed \end{proof}
		
		 We can modify our encoding by setting the neuron potential to be always non-negative and by fixing the threshold  to be 0 as response to an inhibitory stimulation. Recall that a null threshold would make the neuron emit even if its potential is 0.
		Thus, on every firing of the edge $\redge{A}{D}$, if the current sum of weighted inputs $a$ is negative, the threshold $\theta$ is set to 0, otherwise it is set to a $\theta > 0$. This will allow an inhibitory stimulus to produce a rebound spike.

%% file: learning.tex

In this section we examine the \emph{Learning Problem}: \ie how to determine a parameter assignment for a network with a fixed topology and a given input such that a desired output behaviour is displayed. Here we only focus  on the estimation of synaptic weights in a given \SNN; the generalisation of our methodology to other parameters is left for future work.
	
Our analysis takes inspiration from the SpikeProp algorithm \cite{bohte02}; in a similar way, here, the learning process is led by \emph{supervisors}. Differently from the previous section,  each output neuron $\ratuomaton{N}$ is linked to a supervisor instead of an output consumer. Supervisors compare the expected output behaviour  with  the one of the output neuron they are connected to (function $\textsc{Evaluate}(\ratuomaton{N})$ in Algorithm \ref{alg.abp}). Thus either the output neuron behaved consistently or not. In the second case and  in order to instruct the network,  the supervisor  back-propagates \emph{advices} to the output neuron depending on two possible scenarios:
i) the neuron fires a spike, but it was supposed to be quiescent, ii) the neuron remains quiescent, but  it was supposed to fire a spike.
In the first case the supervisor addresses a \emph{should not have fired} message (\snhf) and in the second one a \emph{should have fired} (\shf). Then each output neuron modifies its ingoing synaptic weights and in turn behaves as a supervisor with respect to its predecessors,  back-propagating the proper advice.

The advice back-propagation (\abp), Algorithm \ref{alg.abp}, basically lies  on a depth-first visit of the graph topology of the network.
Let $\ratuomaton{N}_i$ be the $i$-th predecessor of an automaton $\ratuomaton{N}$, then we say that $\ratuomaton{N}_i$ fired, if it emitted a spike  during the current or previous accumulate-fire-wait cycle of $\ratuomaton{N}$. Thus, upon reception of a \shf message, $\ratuomaton{N}$ has to
 \emph{strengthen} the weight of each ingoing \emph{excitatory} synapse  and
		 \emph{weaken} the weight of each ingoing \emph{inhibitory} synapse.
Then, it propagates a \shf advice to each ingoing \emph{excitatory} synapse (\ie an arc with weight greater than 0: $\textsc{Wt} \ge 0$)  corresponding to a neuron which \emph{did not} fire recently ($\neg \textsc{F}(\ratuomaton{N})$ ), and symmetrically  a \snhf advice to each ingoing \emph{inhibitory}   synapse ($\textsc{Wt} < 0$) corresponding to a neuron which fired recently (see Algorithm \ref{alg.abp.shf} for \shf,  and Algorithm \ref{alg.abp.snhf} for the dual case of \snhf). When the graph visit reaches  an input generator,  it will simply ignore any received advice  (because  input sequences should not be affected by the learning process). The learning process ends when all supervisors do not detect any more errors.

\begin{algorithm}[t]
	\caption{ The advice back-propagation algorithm}
	\label{alg.abp}
	\begin{algorithmic}[1]
		\Function{ABP}{}
			\State discovered = $ \emptyset$
			
			\ForAll { $\ratuomaton{N} \in $ Output }
				\If {$\ratuomaton{N} \notin $ discovered }
					\State discovered = discovered $\cup~ \ratuomaton{N} 	$
					\If {\Call{Evaluate}{$\ratuomaton{N}$} = \shf}
						\State \Call{SHF}{$\ratuomaton{N}$}
					\ElsIf {\Call{Evaluate}{$\ratuomaton{N}$} = \snhf}
						\State \Call{SNHF}{$\ratuomaton{N}$}
					\EndIf
				\EndIf
			
			\EndFor
		\EndFunction
	\end{algorithmic}
\end{algorithm}

\begin{algorithm}[t]
	\caption{ Should Have Fired algorithm}
	\label{alg.abp.shf}
	\begin{algorithmic}[1]
		\Procedure{Should-Have-Fired}{$\ratuomaton{N}$}
			\If{$\ratuomaton{N} \in $ discovered $\cup$ Output  }
				\State \Return
			\EndIf
			\State discovered = discovered $\cup~ \ratuomaton{N} 	$
			\ForAll {   $\ratuomaton{M} \in$   \Call{Pred}{$ \ratuomaton{N} $} }			
				\If{$\ratuomaton{M} \notin$ Input }			
				\If{\Call{Wt}{$ \ratuomaton{N}, \ratuomaton{M} $} $\geq 0 ~\wedge  \neg $ \Call{F}{$\ratuomaton{M} $}}
						\State \Call{SHF}{$\ratuomaton{M} $}
				\EndIf
				
				\If{\Call{Wt}{$ \ratuomaton{N}, \ratuomaton{M} $} $ < 0~ \wedge  $ \Call{F}{$\ratuomaton{M} $}}
						\State \Call{SNHF}{$\ratuomaton{M}$}
					
					\EndIf
					
					\EndIf
				\State \Call{Increase-Weight}{$ \ratuomaton{N}, \ratuomaton{M} $}
			\EndFor
			\State \Return
		\EndProcedure
	\end{algorithmic}
\end{algorithm}

\begin{algorithm}[t]
	\caption{Abstract \abp{}: Should \emph{Not} Have Fired advice pseudo-code}
	\label{alg.abp.snhf}
	\begin{algorithmic}[1]
		\Procedure{Should-Not-Have-Fired}{$neuron$} 	
			\If{$\ratuomaton{N} \in $ discovered $\cup$ Output  }
				\State \Return
			\EndIf
			\State discovered = discovered $\cup~ \ratuomaton{N} 	$
	
			\ForAll {   $\ratuomaton{M} \in$   \Call{Predecessors}{$ \ratuomaton{N} $}}			
								\If{$\ratuomaton{M} \notin$ Input }					
				\If{\Call{Wt}{$ \ratuomaton{N}, \ratuomaton{M} $} $\geq 0 ~\wedge   $ \Call{F}{$\ratuomaton{M} $}}
						\State \Call{SNHF}{$\ratuomaton{M} $}
				\EndIf
				
				\If{\Call{Wt}{$ \ratuomaton{N}, \ratuomaton{M} $} $ < 0~ \wedge \neg  $ \Call{F}{$\ratuomaton{M} $}}
						\State \Call{SHF}{$\ratuomaton{M}$}
					
					\EndIf
					\EndIf
				\State \Call{Decrease-Weight}{$ \ratuomaton{N}, \ratuomaton{M} $}
			\EndFor
			\State \Return
		\EndProcedure
	\end{algorithmic}
\end{algorithm}

%
%
%
%
%

\begin{example}[Turning on and off a diamond structure of neurons.]
This example shows how the  \abp algorithm can be used to make a neuron emit at least once in a \SNN having the \emph{diamond} structure shown in \figref{fig.advices.example.diamond.structure}. We assume that $\ratuomaton{N}_1$ is fed by an  input generator $\ratuomaton{I}$ that continuously emits spikes. No neuron in the network is able to emit because all the weights of their input synapses are equal to zero and their thresholds are higher than zero.
We want the network to learn a weight assignment so that  $\ratuomaton{N}_4$ is able to emit, that is, to produce a spike after an initial pause.			
			\begin{figure}[t]
				\centering
					\includegraphics[width=0.8\linewidth]{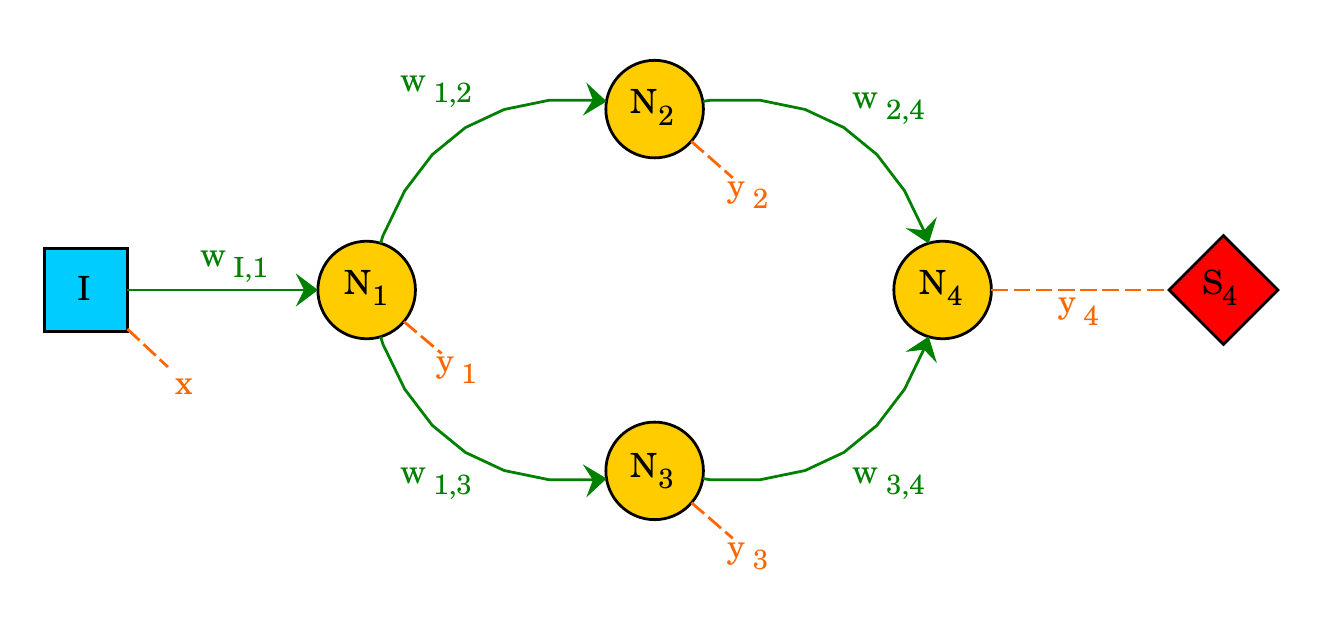}
					\caption{ A neural  network with a diamond structure.
					}
					\label{fig.advices.example.diamond.structure}
	\end{figure}

At the beginning we  expect no activity from neuron $\ratuomaton{N}_4$. As soon as the initial pause is elapsed, we require a spike but, as all weights are equal to zero, no emission can happen.
Thus a \shf advice is back-propagated to neurons $\ratuomaton{N}_2$ and $\ratuomaton{N}_3$ and consequently to $\ratuomaton{N}_1$. The process is then repeated until all weights stabilise and neuron  $\ratuomaton{N}_4$ is able to fire.
\hfill $\diamond$
\end{example}

There are several possibilities on how to realise supervisors and the \abp  algorithm. We propose here two approaches. The first one is model checking oriented and it is based on the idea that supervisors are represented by temporal logic formulae.
The second one is simulation oriented, and the implementation of the algorithm is embedded into the timed automata modelling of the neuron.

\paragraph{ \bf Model-checking-oriented approach.}
\input{second-approach.tex}

\paragraph{\bf Simulation-oriented approach.}
\input{newsimulation.tex}

%% file: second-approach.tex
Such a technique consists in iterating the learning process until a desired CTL property concerning the output of the network is verified. The hypothesis we introduce are the following ones: (i) input generators, standard neurons, and output consumers share a global clock which is never reset and (ii) for each output consumer, there exists a clock measuring the elapsed time since the last received spike. The CTL formula specifying the expected output of the network can only contain predicates relative to the output consumers and the global clock.
At each step of the algorithm, we make an external call to the  model checker to test whether the network satisfies the formula or not. If the formula is verified, the learning process ends; otherwise, the model checker provides a trace as a counterexample. Such a trace is exploited to derive the proper corrective action to be applied to each output neuron, that is, the invocation of either the \shf procedure, or the \snhf procedure previously described (or no procedure).

More in detail, given a \TAN{} representing some \SNN{}, we extend it with a global clock $t_g$ which is never reset and, for each output consumer $\ratuomaton{O}_K$ relative to the output neuron $\ratuomaton{N}_k$, we add a clock $s_k$ measuring the time elapsed since the last spike consumed by $\ratuomaton{O}_k$. Furthermore, let $\pstatesub[O][k]{O}$ be an atomic proposition evaluating to true if the output consumer $\ratuomaton{O}_K$ is in its \textbf{O} location, and let $\pvarsub[O][k]{s_k}$ be an atomic proposition indicating the value of the clock $s_k$ in $\ratuomaton{O}_K$. In order to make it possible to deduce the proper corrective action, we impose the CTL formula describing the expected outcome of the network to be composed by the conjunction of sub-formulae respecting any of the patterns presented in the following. 

\begin{description}
			\item[Precise Firing.] The output neuron $\ratuomaton{N}_k$ fires at time $t$:

				$AF \left(\, t_g = t \wedge \pstatesub[O][k]{O} \,\right)$.

				The violation of such a formula requires the invocation of the \shf procedure.
			
			\item[Weak Quiescence.] The output neuron $\ratuomaton{N}_k$ is quiescent at time $t$:
				
					$AG \left(\, t_g = t \implies \neg \pstatesub[O][k]{O} \,\right)$.
				
				The \snhf procedure is called in case this formula is not satisfied.
			
			\item[Relaxed Firing.] The output neuron $\ratuomaton{N}_k$ fires at least once within the time window $[\, t_1,\, t_2 \,]$:
				
					$AF \left(\, t_1 \leq t_g \leq t_2 \wedge \pstatesub[O][k]{O} \,\right)$.
				
				The violation of such a formula leads to the invocation of the \shf procedure.
			
			\item[Strong Quiescence.] The output neuron $\ratuomaton{N}_k$ is quiescent for the whole duration of the time window $[\, t_1,\, t_2 \,]$:
				
					$AG \left(\, t_1 \leq t_g \leq t_2 \implies \neg \pstatesub[O][k]{O} \,\right)$.
				
				The \snhf procedure is needed in this case.
			
			\item[Precise Periodicity.] The output neuron $\ratuomaton{N}_k$ eventually starts to periodically fire a spike with exact period $P$:
				
					$AF (\, AG (\, \pvarsub[O][k]{s_k} \neq P \implies \neg \pstatesub[O][k]{O} \,) \\
					\wedge
					AG (\, \pstatesub[O][k]{O} \implies \pvarsub[O][k]{s_k} = P \,) \,)$.
				
				If $\ratuomaton{N}_k$ fires a spike while the $s_k$ clock is different than $P$ or it does not fire a spike while the $s_k$ clock  equals $P$, the formula is not satisfied.
				In the former (resp. latter) case, we deduce that the \snhf (resp. \shf) procedure is required.
			\item[Relaxed Periodicity.] The output neuron $\ratuomaton{N}_k$ eventually begins to periodically fire a spike with a period that may vary in $[\, P_{min},\, P_{max} \,]$:
			
				$AF (\, AG (\,  \pvarsub[O][k]{s_k} \notin [\, P_{min},\, P_{max} \,] \implies \\ \qquad \neg \pstatesub[O][k]{O} \,)
				\wedge \\
				AF (\, \pstatesub[O][k]{O} \implies \\
				 P_{min} \leq \pvarsub[O][k]{s_k} \leq P_{max} \,) \,)$.
			
            For the corrective actions, see the previous case.
			
		\end{description}
As for future work, we intend to extend this set of CTL formulae with new formulae concerning the comparison of the output of two or more given neurons.
{Please notice that the Uppaal model-checker only supports a fragment of CTL where the use of nested path quantifiers is not allowed. Another model-checker should be called in order to fully exploit the expressive power of CTL.}

Next we give a couple of examples.
\begin{example}[Diamond network.] \label{ex.diamond.checking}
In this example, we apply the model-checking approach to the diamond network of Figure \ref{fig.advices.example.diamond.structure}. Neuron $\ratuomaton{N}_4$ cannot emit a spike with the initial weights and parameters, which are:
\begin{center}
	\begin{tabular}{|c|c|c|c|c|}
		\hline
		$w_{0,1}$ & $w_{1,2}$ & $w_{1,3}$ & $w_{2,4}$ & $w_{3,4}$ \\
		\hline
		$0.1$ & $0.1$ & $0.1$ & $0.1$ & $0.1$ \\
		\hline
	\end{tabular}
\end{center}

\begin{center}
	\begin{tabular}{c|c|c|c|c}
	Neuron & T & $\theta$ & $\tau$ & $\lambda$\\
	\hline
	$\ratuomaton{N}_1$ & $2$ & $0.35$ & $3$ & $7/9$\\
	$\ratuomaton{N}_2$ & $2$ & $0.35$ & $3$ & $7/9$\\
	$\ratuomaton{N}_3$ & $2$ & $0.35$ & $3$ & $7/9$\\
	$\ratuomaton{N}_4$ & $2$ & $0.55$ & $3$ & $1/2$\\
	\end{tabular}
\end{center}

We expect  $\ratuomaton{N}_4$ to spike every 20 time units. After three cycles of the algorithm (\ie three checks of the formula and modifications of weights), we reach the following weight assignment:

\begin{center}
	\begin{tabular}{|c|c|c|c|c|}
		\hline
		$w_{0,1}$ & $w_{1,2}$ & $w_{1,3}$ & $w_{2,4}$ & $w_{3,4}$ \\
		\hline
		$0.3$ & $0.3$ & $0.3$ & $0.3$ & $0.3$ \\
		\hline
	\end{tabular}
\end{center}

~ \hfill $\diamond$
\end{example}
		
\begin{example}[Mutual inhibition networks.]
			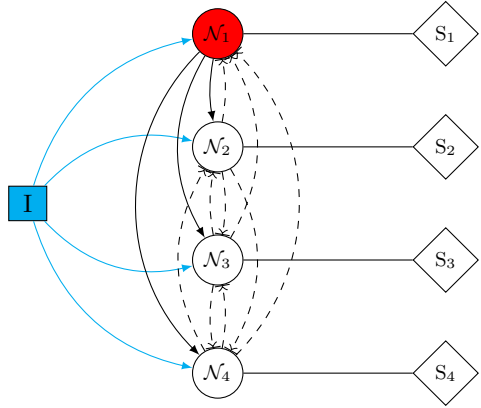
\begin{figure}[t]
				\centering

\begin{tikzpicture}
		\node[draw, minimum width =0.5cm, fill=cyan] (I) at (1.5,0.75) { I } ;
		\node[draw,circle, scale = 0.8, fill=red] (a) at (4,3) {$\ratuomaton{N}_1$};
		\node[draw,circle, scale = 0.8] (b) at (4,1.5) {$\ratuomaton{N}_2$};
		\node[draw,circle, scale = 0.8] (c) at (4,0) {$\ratuomaton{N}_3$};
		\node[draw,circle, scale = 0.8] (d) at (4,-1.5) {$\ratuomaton{N}_4$};

		\node[draw,diamond, scale = 0.8] (o1) at (7, 3) {S$_1$};
		\node[draw,diamond, scale = 0.8] (o2) at (7, 1.5) {S$_2$};
		\node[draw,diamond, scale = 0.8] (o3) at (7, 0) {S$_3$};
		\node[draw,diamond, scale = 0.8] (o4) at (7, -1.5) {S$_4$};

		\draw[->, >=latex, cyan] (I)  to[bend left = 30] (a);
		\draw[->, >=latex, cyan] (I)  to[bend left = 30] (b);
		\draw[->, >=latex, cyan] (I)  to[bend right = 30] (c);
		\draw[->, >=latex, cyan] (I)  to[bend right = 30] (d);

		\draw[->, >=latex] (a) to[bend right = 10] (b);
		\draw[->, >=latex] (a) to[bend right = 30] (c);
		\draw[->, >=latex] (a) to[bend right = 45] (d);
		
		\draw[->|, dashed] (b) to[bend right = 10] (a);
		\draw[->|, dashed] (b) to[bend left = 10] (c);
		\draw[->|, dashed] (b) to[bend left = 30] (d);

		\draw[->|, dashed] (c) to[bend right = 30] (a);
		\draw[->|, dashed] (c) to[bend left = 10] (b);
		\draw[->|, dashed] (c) to[bend right = 10] (d);

		\draw[->|, dashed] (d) to[bend right = 45] (a);
		\draw[->|, dashed] (d) to[bend left = 30] (b);
		\draw[->|, dashed] (d) to[bend right = 10] (c);

		\draw(a) -- (o1);
		\draw(b) -- (o2);
		\draw(c) -- (o3);
		\draw(d) -- (o4);
	\end{tikzpicture}					
					
%
%
%
%
%
%
%
%
%

					\caption{
						We denote neurons by $\ratuomaton{N}_i$.
					The network is fed by an input generator $\ratuomaton{I}$ and the learning process is led by the supervisors $\ratuomaton{S}_i$.
Dotted (resp. continuous) edges stand for inhibitions (resp. activations).  
					}
					\label{fig.advices.example.winner.structure}
	\end{figure}	
In this example we focus on mutual inhibition networks, where the constituent neurons inhibit each other neuron's activity.
These networks belong to the set of Control Path Generators (CPGs), which are known for their capability to produce
rhythmic patterns of neural activity without receiving rhythmic inputs \cite{Ijs08}. CPGs
underlie many fundamental rhythmic activities such as digesting,
breathing, and chewing. They are also crucial building
blocks for the locomotor neural circuits both in invertebrate
and vertebrate animals.
It has been observed that, for suitable parameter values, mutual inhibition networks present a behaviour of the kind "winner takes all", that is,
at a certain time one neuron becomes (and stays) activated and the other ones are inhibited \cite{demaria16}.

We consider a mutual inhibition network of four neurons, as shown in Figure \ref{fig.advices.example.winner.structure}.
This example, although being small, it is not trivial as it features inhibitor and excitatory edges as well as cycles.

We look for synaptical weights such that the "winner takes all" behaviour is displayed. We assume each neuron to be fed by an  input generator $\ratuomaton{I}$ that continuously emits spikes. At the beginning, all the neurons have the same parameters (that is, firing threshold, remaining coefficient, accumulation period, and refractory period), and the weight of excitatory (resp. inhibitory) edges is set to 1 (resp. -1). We use the \abp algorithm to learn a weight assignment so that the first neuron is the winner. More precisely, we find a weight assignment so that, whatever the chosen path in the corresponding automata network is, the network stabilises when the global clock $t_g$ equals 70. The weight of the edges from the input generator $\ratuomaton{I}$ to the four neurons equals 0.041.
The weight of the edges inhibiting $\ratuomaton{N}_1$ (resp. $\ratuomaton{N}_2$, $\ratuomaton{N}_3$, and $\ratuomaton{N}_4$) is -0.719 (resp. -0.817).

~ \hfill $\diamond$
\end{example}

%% file: newsimulation.tex

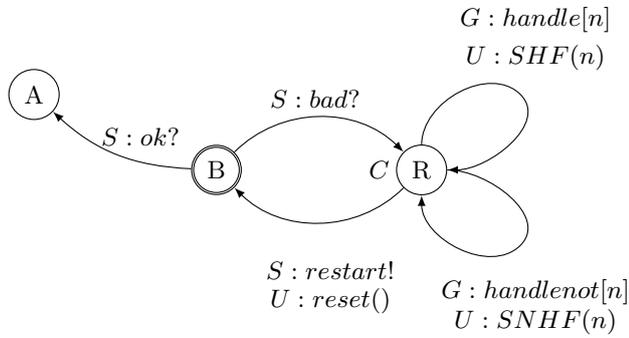
\begin{figure}[t]
	
	\begin{center}
	\begin{tikzpicture}
		\node[draw,circle] (B) at (0,0) {B};
		\node[draw, circle, minimum width = 0.6cm] (B2) at (0,0){};
		\node[draw,circle] (R) at (2.7,0) {R};
		\node[draw,circle] (A) at (-2.4,1) {A};

		\draw(2.4,0) node[left]{$C$};
		
		\draw[->, >=latex](B) to[bend left = 45](R);
		\draw(1.3, 0.7) node[above]{$S : bad?$};

		\draw[->, >=latex](R) to[bend left = 45](B);
		\draw(1.5, -1.1) node[below]{$S : restart!$};
		\draw(1.5, -1.45) node[below]{$U : reset()$};

		\draw[->, >=latex](B) to[bend left = 20](A);
		\draw(-1, 0.2) node[above]{$S : ok?$};

		\draw[->, >=latex](R) to[out = 90, in = 135] (4,1) to[out = -45, in = 0] (R);
		\draw(4.2,1.7) node[above] {$G : handle[n]$};
		\draw(4.2,1.2) node[above] {$U : \shf(n)$};

		\draw[->, >=latex](R) to[out = 0, in = 45] (4,-1) to[out = 225, in = -90] (R);
		\draw(4.2, -1.9) node[above] {$G :handlenot[n]$};
		\draw(4.2,-2.3) node[above] {$U : \snhf(n)$};
		
	\end{tikzpicture} 	
	\end{center}
	\caption{The automata responsible of the  $ABP-alg$}\label{fig:algo}
\end{figure}

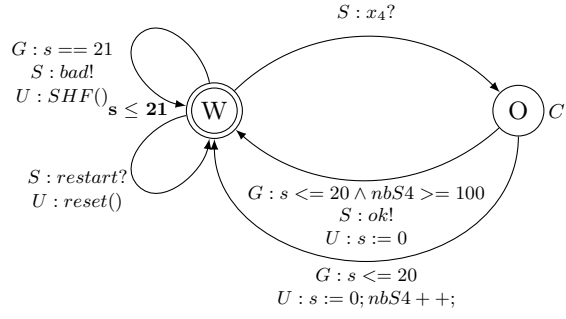
\begin{figure}[t]
		\begin{center}
	\begin{tikzpicture}	
		\node[draw,circle] (W) at (0,0) {W} ;
		\node[draw, circle, minimum width = 0.6cm](W2) at (0,0) {};
		\node[draw,circle] (O) at (4,0) {O};
		\draw(4.3,0) node[right, scale = 0.8] {$C$};

		\draw[->, >=latex] (W) to[bend left = 45] (O); 	
		\draw (2, 1.5) node[below, scale = 0.75]{$S : x_4?$} ;


		\draw[->, >=latex] (O) to[bend left = 42] (W);
		\draw(2, -1.2) node[below, scale = 0.75]{$S : ok!$} ;
		\draw(2, -0.9) node[below, scale = 0.75]{$G : s <= 20 \land nbS4 >= 100 $} ;
		\draw(2, -1.5) node[below, scale = 0.75]{$U :  s:=0 $} ;

		\draw[->, >=latex] (O) to [out = 270, in = 0] (2,-2) to[out = 180 , in = 270] (W);
		\draw(2, -2) node[below, scale = 0.75]{$G : s <= 20 $} ;
		\draw(2, -2.3) node[below, scale = 0.75]{$U :  s:=0;  nbS4++; $} ;

		\draw[->, >=latex] (W) to[out = 100, in = 45] (-1,1) to[out = 225, in = 170] (W);
		\draw(-2,1) node[below, scale = 0.75]{$G : s == 21$};
		\draw(-2,0.7) node[below, scale = 0.75]{$S : bad!$};
		\draw(-2, 0.4) node[below, scale = 0.75]{$U : \shf()$};

		\draw[->, >=latex] (W) to[out = 190, in = 135] (-1,-1) to[out = 315, in = 260] (W);
		\draw(-1.8, -0.7) node[below, scale = 0.75]{$S : restart?$};
		\draw(-1.8, -1) node[below, scale = 0.75]{$U : reset()$};
		\draw(-1,0) node[scale = 0.75] {\textbf {s} $\leq$ \textbf{21}};
		\end{tikzpicture} \linebreak
	\end{center}
	\caption{Example of an output consumer in the simulation approach for the diamond structure}\label{fig:consumersnew}
\end{figure}

In this second approach,  parameters  are modified during the simulation of the network.
This entails that the encoding of neurons needs to be adjusted in order to take care of the adaptation of such parameters. Algorithm ABP is realised by a dedicated automaton $ABP-alg$, that is depicted in Figure \ref{fig:algo}, and the role of supervisor is given to output consumers, that are modified as in Figure \ref{fig:consumersnew}.

The idea is that, according to the function \textsc{Evaluate}, if the corresponding output neuron misbehaves, then its output consumer sets whether it has to be treated according to the \shf or the \snhf  function. Furthermore, it signals to the $ABP-alg$ through the message $bad!$ that some adjustments on the network have to be done. Then the $ABP-alg$ automaton takes the lead and it recursively  applies the function \shf or \snhf (this is achieved by setting a proper variable in a vector named $handle$) on the predecessors of the output neuron.
Once there is no more neuron to whom the algorithm should be applied (for instance all neurons in the current run have been visited), the simulation is restarted in the network with the new parameters.
If the output consumer does not recognise any misbehaviour, than it sends an $ok!$ message to the $ABP-alg$ automaton, that in turn moves to an accepting state $A$.

More formally:
\begin{definition}[Output consumer for the simulation approach]
An output consumer for the simulation approach is a timed automaton $${\mathcal{N}} = (L,\rstate{A},X,Var,\Sigma,\arcs,\inv)$$ with:

\begin{itemize}
				 \item $L = \{ \rstate{W}, \rstate{O}\}$  with $\rstate{O}$ committed,

				\item $X=\{s\}$
				\item $Var=\{ nb, handle[N], handlenot[N]\}$
				\item $\Sigma = \{ x_i \mid x_i \text{ is an output neuron}\} \cup \{ bad, restart, ok\}$,
				\item $\arcs=\{ (\rstate{W}, s<T_{min}, bad!, \{\shf()\}, \rstate{W}), \\
				 (\rstate{W},  , restart?, \{s:=0, nb:=0\}, \rstate{W}), \\
				 (\rstate{W},  ,x_i?, ,\rstate{0}),
				 (\rstate{O},  s>T_{max}, bad!, \{\snhf()\},\rstate{W}), \\
				 (\rstate{O},  , restart?, \{s:=0, nb:=0\}, \rstate{O}), \\
			(\rstate{O},  good\_pattern, ok!, \{s:=0\},\rstate{W}),\\
			(\rstate{O},  good\_not\_finished, , \{s:=0\},\rstate{W})		
					\}$  ;
			  \item $\inv(\rstate{W}) = s \leq T$, $\inv(\rstate{O}) = \true$
			\end{itemize}
where the functions \shf and \snhf modify the global variables $handle_i$  and $handlenot_i$ respectively, that are used in the $ABP-alg$ automaton.
\end{definition}

Notice that the precise definition (for instance the value of  parameters  $good\_pattern$,
$good\_not\_finished$, $T_{min}$, $T_{max}$) of the output consumer depends on the expected behaviour of the supervisor. As an example, in Figure \ref{fig:consumersnew}, that depicts the output consumer for the diamond network,  we expect the output neuron to emit a spike within each window of 20 seconds for at least 100 times.

More in detail, the cycle from $\rstate{W}$ to $\rstate{W}$ is taken whenever a spike has not been sent before $T$ time units. Thus the automaton sends to the $ABP-alg$ automaton a $bad$ message (signalling that an adjustment to the network should take place), and the update part handles the fact that we should perform the \shf algorithm (a variable corresponding to the concerned output neuron is set to $\true$ in the array $handle$).	
From the same state, whenever the message $restart$ is received, all the variables of the automaton  are reset to 0.
When a spike from the output neuron $x$ is received, the output consumer moves to the state $\rstate{O}$. In this state, if the spike was received too late (and similarly as in the previous case), a $bad$ message is sent to the $ABP-alg$. Otherwise, if everything was received on time and if the expected pattern has been completely verified, then an $ok$ message is sent and the automaton moves to the initial state.

In the following, we give the formal definition of the $ABP-alg$ automaton:
\begin{definition}[$ABP-alg$ automaton]
\begin{itemize}
				\item $L = \{ \rstate{A}, \rstate{B}, \rstate{R}\}$  with $\rstate{R}$ committed,
				\item $X=\emptyset$
				\item $Var=\{ handle[N], handlenot[N] \}$
				\item $\Sigma = \{ bad, ok, restart \}$,
				\item $\arcs=\{
				        (\rstate{B}, , ok?, ,\rstate{A}), (\rstate{B}, , bad?, ,\rstate{R}) \\
				        ( \rstate{R}, handle[n] ,, \shf(n) ,  \rstate{R})  \\
				         ( \rstate{R}, handlenot[n] ,, \snhf(n) ,  \rstate{R})\\
				         ( \rstate{R}, finished?() , restart! , reset() ,  \rstate{B})			        \}$  ;
			  \item $ \inv(\rstate{A}) = \true, \inv(\rstate{B}) = \true, \inv(\rstate{R}) = \true$.
			\end{itemize}

\end{definition}
The arc from the state $\rstate{B}$ to the accepting state $\rstate{A}$ is taken whenever the Output consumer  has finished the analysis of its pattern. Conversely, the arc from the state $\rstate{B}$ to $\rstate{R}$ is adopted when one of the output neurons has misbehaved (signalled by the reception of the message $bad$). In the committed state $\rstate{R}$, the parameters of the neural network are changed accordingly to the Algorithms \ref{alg.abp.shf} and \ref{alg.abp.shf}. The arrays $handle$ (respectively $handlenot$) have the information on the neurons to which the Algorithm for \shf (respectively \snhf) has to be applied. Once the cycle of updates finishes (checked through the function $finished?()$), the simulation is restarted by broadcasting a message $restart$ to all the neurons and the output consumer.

Last, we give the definition of the changes induced in the standard neuron. As the update of the neuron parameters is done at the level of $ABP-alg$, the only change concerns the treatment of the signal $restart$. To this aim an arc handling the reception of the message is added to the states $\rstate{W}$ and $\rstate{A}$.
\begin{definition}[ Standard Neuron for the simulation approach]
			Given a neuron $v=(\theta, \tau, \lambda, p, y)$ with $m$ input synapses, its encoding into timed automata is  ${\mathcal{N}} = (L,\rstate{A},X,Var,\Sigma,\arcs,\inv)$ with:
		
			\begin{itemize}
				\item $L = \{\rstate{A}, \rstate{W}, \rstate{D}\}$  with $\rstate{D}$ committed,
				\item $X=\{t\}$
				\item $Var=\{ p,a\}$
				\item $\Sigma = \{ x_i \mid i \in [1..m] \} \cup \{ y\}$,
				\item $\arcs=\{ (\rstate{A}, t\leq T, x_i?, \{a:=a+w_i\}, \rstate{A}) \mid i \in [1..m]\}
					\cup \{(\rstate{A}, t = T, ~, \{p:=a+\lfloor \lambda p \rfloor \}, \rstate{D}), \\
					(\rstate{D}, p < \theta, ~, \{a:=0\}, \rstate{A}),
					(\rstate{D}, p \geq \theta, y!, ~, \rstate{W}),\\
					(\rstate{W}, t=\tau , ~, \{a:=0, t:=0, p:=0\}, \rstate{A}),\\
					(\rstate{A}, , restart?, \{a"=0, t:=0, p:=0 \} \rstate{A}), \\
					(\rstate{W}, , restart?, \{a"=0, t:=0, p:=0 \} \rstate{A})
					\}$  ;
			  \item $\inv(\rstate{A}) = t \leq T, \inv(\rstate{W}) = t \leq \tau, \inv(\rstate{D}) = \true$.
			\end{itemize}

\end{definition}

%

	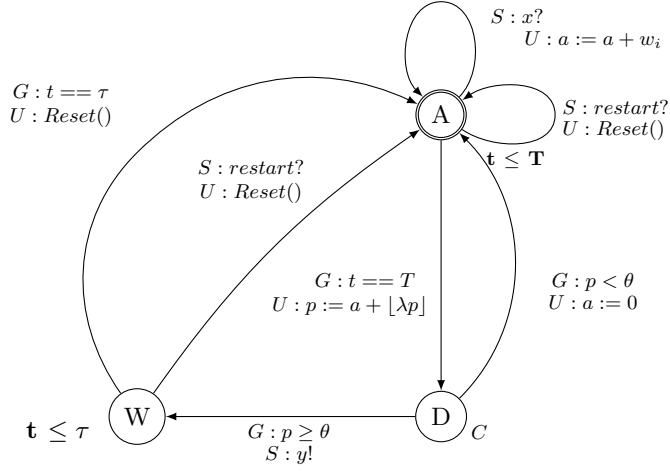
\begin{figure}[t]
	\newcommand{\snb}{0.8}
		\caption{Example of a neuron in the simulation approach for the diamond network }
		\begin{center}
		\begin{tikzpicture}[scale= 1]
			\node[draw,circle] (A) at (4,0) {A} ;
			\node[draw, circle, minimum width = 0.6cm](A2) at (4,0) {};
			\node[draw,circle] (D) at (4,-4) {D};
			\node[draw,circle] (W) at (0,-4) {W};
			
			\draw(4.5,-0.6) node[right,scale =\snb ] {\textbf{t $\leq$ T}};
			\draw(4.3,-4.2) node[right,scale =\snb ] {$C$};
			\draw(-0.5,-4.2) node[left]{\textbf{t $\leq\tau$}};

			\draw[->, >=latex] (A) -- (D);
			\draw (3, -2) node[below,scale =\snb ]{$G : t==T$} ;
			\draw (2.8, -2.3) node[below,scale =\snb ]{$U : p:= a + \lfloor \lambda p \rfloor$};

			\draw[->, >=latex] (D) -- (W); 	
			\draw (2, -4) node[below,scale =\snb ]{$G : p\geq\theta$} ;
			\draw (2, -4.3) node[below,scale =\snb ]{$S : y!$} ;
			
			\draw[->, >=latex] (W) to[out = 125, in = 225] (0,-0.5) to[out = 45, in = 155]  (A);
			\draw (-1, 0.5) node[below,scale =\snb ]{$G : t==\tau$} ;
			\draw (-1, 0.2) node[below,scale =\snb ]{$U : Reset()$} ;

			\draw[->, >=latex] (A) to[out = 45, in = 0] (4,1.5) to[out = 180, in = 135 ] (A);
			\draw (5, 1.5) node[below,scale = \snb ]{$S : x?$} ;
			\draw (6, 1.2 ) node[below,scale = \snb ]{$U : a:=a+w_i$} ;

			\draw[->, >=latex] (D) to[bend right = 45] (A);
			\draw (6, -2) node[below,scale = \snb ]{$G : p<\theta$} ;
			\draw (6, -2.3) node[below,scale = \snb ]{$U : a := 0$};

			\draw[->, >=latex] (W) to[bend left = 10 ] (A);
			\draw(1.5,-0.5)node[below, scale = \snb]{$S : restart?$};
			\draw(1.5,-0.8)node[below, scale = \snb]{$U : Reset()$};

			\draw[->, >=latex] (A) to[out = 325 , in = 270] (5.5,0) to[out =90 , in =35](A) ;
			\draw(5.5,0.1) node[right, scale =	\snb] {$S : restart?$};
			\draw(5.5,-0.2) node[right, scale =	\snb]{$U : Reset()$};
		\end{tikzpicture}
		\end{center}
		\label{autoNeurones}

	\end{figure}

Notice that the algorithm can be refined by setting some priorities on the neurons. For instance, if any additional information is known in advance on the behaviour of a specific neuron, its actions can be constrained and the corrective operations could be ignored.
In some cases, the type of synapse (excitatory or inhibitory) can also be set, disallowing the possibility of changing its nature (\eg from inhibitory to excitatory).

\begin{example}[Diamond network]
In this example, we apply the simulation-oriented approach on the diamond network on Figure \ref{fig.advices.example.diamond.structure}.
We take the same parameters and aim as in Example \ref{ex.diamond.checking}.
%

After the simulation, we obtain the following resulting weights :
\begin{center}
	\begin{tabular}{|c|c|c|c|c|}
		\hline
		$w_{0,1}$ & $w_{1,2}$ & $w_{1,3}$ & $w_{2,4}$ & $w_{3,4}$ \\
		\hline
		$0.2$ & $0.3$ & $0.3$ & $0.3$ & $0.3$ \\
		\hline
	\end{tabular}
\end{center}

We obtain these weights when the global clock equals 42 and after only 2 sends of the message $bad$ (meaning that the neuronal network is stabilising in only two applications of the \abp algorithm).
%
%
Note that the resulting weights for the model-checking-oriented approach, seen in the precedent example, are different. Indeed, different sets of weights can give the same behaviour.

~ \hfill $\diamond$
\end{example}

\begin{example}[Mutual inhibition network ]
\begin{figure}[t]
				\centering
\begin{tikzpicture}
		\node[draw, minimum width =0.5cm, fill=cyan] (I) at (1.5,0) { $\ratuomaton{I}$ } ;
		\node[draw,circle, fill=red, scale = 0.8] (a) at (4,2) {$\ratuomaton{N}_1$};
		\node[draw,circle, scale = 0.8] (b) at (4,0) {$\ratuomaton{N}_2$};
		\node[draw,circle, scale = 0.8] (c) at (4,-2) {$\ratuomaton{N}_3$};

		\node[draw,diamond, scale = 0.8] (o1) at (7, 2) {$\ratuomaton{S}_1$};
		\node[draw,diamond, scale = 0.8] (o2) at (7, 0) {$\ratuomaton{S}_2$};
		\node[draw,diamond, scale = 0.8] (o3) at (7, -2) {$\ratuomaton{S}_3$};

		\draw[->, >=latex, cyan] (I)  to[bend left = 30] (a);
		\draw[->, >=latex, cyan] (I)  -- (b);
		\draw[->, >=latex, cyan] (I)  to[bend right = 30] (c);

		\draw[->|, dashed] (a) to[bend right = 10] (b);
		\draw[->|, dashed] (a) to[bend right = 30] (c);

		\draw[->|, dashed] (b) to[bend right = 10] (a);
		\draw[->|, dashed] (b) to[bend left = 10] (c);

		\draw[->, >=latex] (c) to[bend right = 30] (a);
		\draw[->, >=latex] (c) to[bend left = 10] (b);

		\draw(a) -- (o1);
		\draw(b) -- (o2);
		\draw(c) -- (o3);
	\end{tikzpicture}
	\caption{ A second neural network with "winner takes all" behaviour }\label{fig.advices.example.winner2.structure}
	\end{figure}
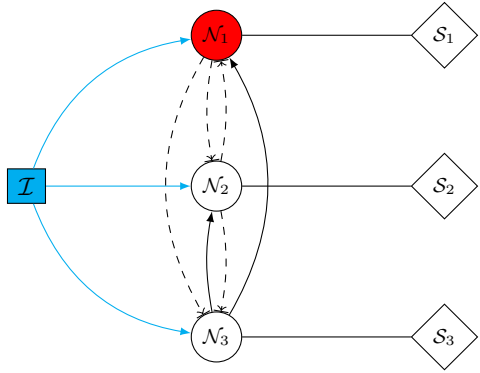	
This example shows the weight estimation for the network of  Figure \ref{fig.advices.example.winner2.structure} with the simulation approach. As before, we require a ``winner takes all" behaviour where $\ratuomaton{N}_1$ is the winner. We expect that  $\ratuomaton{N}_1$ spikes every 10 time units at most, and the neurons $\ratuomaton{N}_2$ and $\ratuomaton{N}_3$ do not spike. We take a network with the following initial parameters: 

 \begin{center}
  \begin{tabular}{c|c|c|c|c}
	Neuron & T & $\theta$ & $\tau$ & $\lambda$\\
	\hline
	$\ratuomaton{N}_1$  & $2$ & $0.75$ & $3$ & $1/2$ \\
	$\ratuomaton{N}_2$  & $2$ & $0.75$ & $3$ & $1/2$ \\
	$\ratuomaton{N}_3$  & $2$ & $0.75$ & $3$ & $1/2$ \\
  \end{tabular}
 \end{center}

 \begin{center}
 \begin{tabular}{c|c|c|c|c|}
 	$w_{x,y}$& $\ratuomaton{I}$ & $\ratuomaton{N}_1$ & $\ratuomaton{N}_2$ & $\ratuomaton{N}_3$ \\
	\hline
 	$\ratuomaton{I}$	& 0 & 0.1 & 0.1 	& 0.1 	\\	 	
 	\hline	
 	$\ratuomaton{N}_1$	& 0 & 0  	& -0.1 	& -0.1 	\\				
	\hline
 	$\ratuomaton{N}_2$	& 0 & -0.1 & 0 	& -0.1 	\\			
	\hline
 	$\ratuomaton{N}_3$	& 0 & 0.1 & 0.1 	& 0 	\\	
  \end{tabular}
 \end{center}

And we obtain the following weights in 5 executions of the \abp algorithm:
 \begin{center}
  \begin{tabular}{c|c|c|c|c|}
 	$w_{x,y}$& $\ratuomaton{I}$ & $\ratuomaton{N}_1$ & $\ratuomaton{N}_2$ & $\ratuomaton{N}_3$ \\
	\hline
 	$\ratuomaton{I}$	& 0 & 0.6 & 0.1 	& 0.1 	\\	 	
 	\hline	
 	$\ratuomaton{N}_1$	& 0 & 0  	& -0.1 	& -0.1 	\\				
	\hline
 	$\ratuomaton{N}_2$	& 0 & -0.1 & 0 	& -0.1 	\\			
	\hline
 	$\ratuomaton{N}_3$	& 0 & 0.6 & 0.1 	& 0 	\\	
  \end{tabular}
 \end{center}

%
%
%
~ \hfill $\diamond$
\end{example}
Notice that, as the application of the \abp algorithm in the  simulation approach is non-deterministic, the parameters we found may depend on the precise execution and several solutions are therefore possible. 
The complete encoding of the examples shown here can be found at \cite{page_laetitia}.

%% file: related.tex
To the best of our knowledge, there are few attempts of giving formal models for \lif. Apart from the already discussed approach of \cite{demaria16}, where the authors model and verify \lif networks thanks to the synchronous language Lustre, the closest related work we are aware of is \cite{ciobanu16}. In this work, the authors propose a mapping of spiking neural P systems into timed automata. The modelling is substantially different from ours. They consider neurons as static objects and the dynamics is given in terms of evolution rules while for us the dynamics is intrinsic to the modelling of the neuron. This, for instance, entails that inhibitions are not just negative weights as in our case, but are represented as \emph{forgetting rules}. On top of this, the notion of time is also different: while they consider  durations in terms of number of applied rules, we have an explicit notion of duration given in terms of accumulation and refractory period.

As far as our parameter learning approach is concerned, we borrow inspiration from the SpikeProp rule \cite{bohte02}, a variant of the well known back-propagation algorithm \cite{rumelhart88} used for supervised learning in second generation learning. The SpikeProp rule deals with multi-layered cycle-free \SNN{s} and aims at training networks to produce a given output sequence for each class of input sequences.
The main difference with respect to our approach is that we are considering here a discrete model and our networks are not multi-layered.
We also rest on Hebb's learning rule \cite{hebb1950} and its time-dependent generalisation rule, the spike timing dependent plasticity (STDP) rule \cite{sjostrom2010}, which aims at adjusting the synaptical weights of a network according to the time occurrences of input and output spikes of neurons. It acts locally, with respect to each neuron, \ie no prior assumption on the network topology is required in order to compute the weight variations for some neuron input synapses.
Differently from the \stdp, our approach takes into account not only recent spikes but also some external feedback (\emph{advices})  in order to determine which weights should be modified and whether they must increase or decrease. Moreover, we do not prevent excitatory synapses from becoming inhibitory (or vice versa), which is usually a constraint for \stdp implementations. A general overview on \SNN learning approaches and open problems in this context can be found in \cite{gruning14}.

%% file: conclusion.tex
In this paper we formalised the \lif  model of \SNN{s} via \TAN{s}.
We have a complete implementation of the proposed  model and examples via the tool \Uppaal, that can be found at the pages \cite{web} and \cite{page_laetitia}. 
In our modeling framework, information processing is based on the precise
timing of spike emissions rather than the average numbers of spikes
in a given time window. Timed automata turned out to be very suited to model \SNN{s}, allowing us to take into account  time-related aspects, such as the exact spike occurrence times and the refractory period, a lapse of time immediately following each spike emission, when the neuron emission capability is reduced.

In this work, we exploited model checking to automatically validate our automaton-based mapping of the \lif model according to a number of behaviours (i.e., typical
responses to an input pattern) the \lif model should be able to
reproduce, namely tonic spiking, excitability, and integrator.
Formal methods of computer science turned out to be an effective tool to validate our modelling approach.
As for future work concerning the modeling aspects, we plan to provide analogous formalisations for
more complex spiking neuron models, such as the theta-neuron model \cite{ermentrout86}
or the Izhikevich one \cite{izhikevich03}. We also intend to extend our model to include propagation
delays, which are considered important within the scope
of spiking neural networks \cite{paugam-moisy2012}.
Our extension is intended to add  suitable states and clocks to model synapses. We also plan to perform
a robustness analysis of the obtained model, in order to detect which neuron parameters influence most the verification
of some wished temporal properties.

As key contribution, we proposed a novel technique to infer the synaptical weights of \SNN{s}.
At this aim, we adapted machine learning techniques to bio-inspired models, which makes our work original and
complementary with respect to the main international projects aiming at understanding the human brain,
such as the Human Brain Project \cite{hbp}, which mainly relies on large-scale simulations.

For our learning approach, we have focussed on a simplified type of supervisors: each supervisor describes the output of a single neuron in isolation from the other ones. Nonetheless, notice that the back-propagation algorithm  is still valid for more complex scenarios that specify and compare the behaviour of groups of neurons. As for future work, we intend to formalise more sophisticated supervisors, allowing to compare the output of several neurons.
Moreover, to refine our learning algorithm, we could exploit results coming from the gene regulatory network domain, where  a link between the topology of the network and its dynamical behaviour is established \cite{adrien2010}.

As a last step, we plan to generalise our technique in order to be able to infer not only synaptical weights but also other parameters, such as the leak factor or the firing threshold.